\documentclass[
  prb,
  superscriptaddress,
  10 pt, twocolumn,
 amsmath,amssymb
]{revtex4-2}

\usepackage{graphicx}
\usepackage{bm}

\begin{document}

\title{The Structure and Dispersion of Exciton-Trion-Polaritons in Two-Dimensional Materials: Experiments and Theory}

\author{Okan Koksal}
\affiliation{ 
	School of Electrical and Computer Engineering, Cornell University, Ithaca, NY 14853, USA
}%
\author{Minwoo Jung}%
\affiliation{ 
	Department of Physics, Cornell University, Ithaca, NY 14853, USA
}%

\author{Christina Manolatou}
\affiliation{ 
	School of Electrical and Computer Engineering, Cornell University, Ithaca, NY 14853, USA
}%

\author{A. Nick Vamivakas}
\affiliation{ Institute of Optics, University of Rochester, Rochester, NY, USA}

\author{Gennady Shvets}%
\affiliation{ 
	School of Applied and Engineering Physics, Cornell University, Ithaca, NY 14853, USA
}%
\author{Farhan Rana}
\affiliation{ 
	School of Electrical and Computer Engineering, Cornell University, Ithaca, NY 14853, USA
}%

\email{ok74@cornell.edu}

\begin{abstract}
  The nature of trions and their interaction with light has remained a puzzle. The composition and dispersion of polaritons involving trions provide insights into this puzzle. Trions and excitons in doped two-dimensional (2D) materials are not independent excitations but are strongly coupled as a result of Coulomb interactions. When excitons in doped 2D materials are also strongly coupled with light inside an optical waveguide, the resulting polariton states are coherent superpositions of exciton, trion, and photon states. We realize these exciton-trion-polaritons by coupling an electron-doped monolayer of two-dimensional material MoSe$_2$ to the optical mode in a photonic crystal waveguide. Our theoretical model, based on a many-body description of these polaritons, reproduces the measured polariton energy band dispersion and Rabi splittings with excellent accuracy. Our work sheds light on the structure of trion states in 2D matrials and also on the indirect mechanism by which they interact with light.         
\end{abstract}

\maketitle

\section{\label{sec:level1}Introduction}
Ever since the first theoretical description of trions~\cite{trionfirst}, and their subsequent experimental observations~\cite{trfirstexp,gaastrion}, the nature of the trion state in doped semiconductors has remained somewhat of a mystery. The conventional and widely accepted description of a trion as a fermionic bound state of an exciton and a free-charge carrier, while successful in predicting the binding energies of trions (at least at small doping densities)~\cite{Suris01,Ctrion0,Berk13,Rana14,Urba17}, is incompatible with the notion of a trion getting created directly with the absorption of a photon, which is a boson. Perhaps nowhere else is this problem more acute than in the case of polaritons involving trions because quantum states involving a superposition of a fermion and a boson are prohibited by quantum superselection rules.

Various bosonic trion-hole states (assuming an electron-doped material) have been proposed in the literature as an alternative to the fermionic trion states~\cite{EsserTrion,Ctrion,Crecent,Ctrion2,Nonlineartr20}. These trion-hole states, in addition to the trion, consist of a hole in the Fermi sea. This Fermi hole is created when an electron is scattered out of the Fermi sea by a photogenerated exciton to form a trion~\cite{EsserTrion,Ctrion,Crecent,Ctrion2,Nonlineartr20}. Although trion-hole states enjoyed some success in explaining the optical absorption spectra of 2D materials, they fail to describe coherent polaritons involving trions and lead to incorrect results. We emphasize here that this inability to describe polaritons is not related to a lack of accuracy but to fundamental errors in the description of the trion states and their interaction with light, and these errors make polaritons involving trions impossible. A brief technical discussion of various trion states proposed in the literature, and their shortcomings in explaining coherent polaritons involving trions, is included in the Supplementary Material (SM).

Several recent experimental works have reported signatures of polaritons involving trions in two-dimensional (2D) materials~\cite{nickTrion,Emmanuele2020,Cuadra2018,Dufferwiel2017}. These experimental observations beg the question how trions are able to form polaritons and whether the existing descriptions of trions can explain these polaritons.    

\begin{figure}[t]
	\includegraphics[width=0.95\columnwidth]{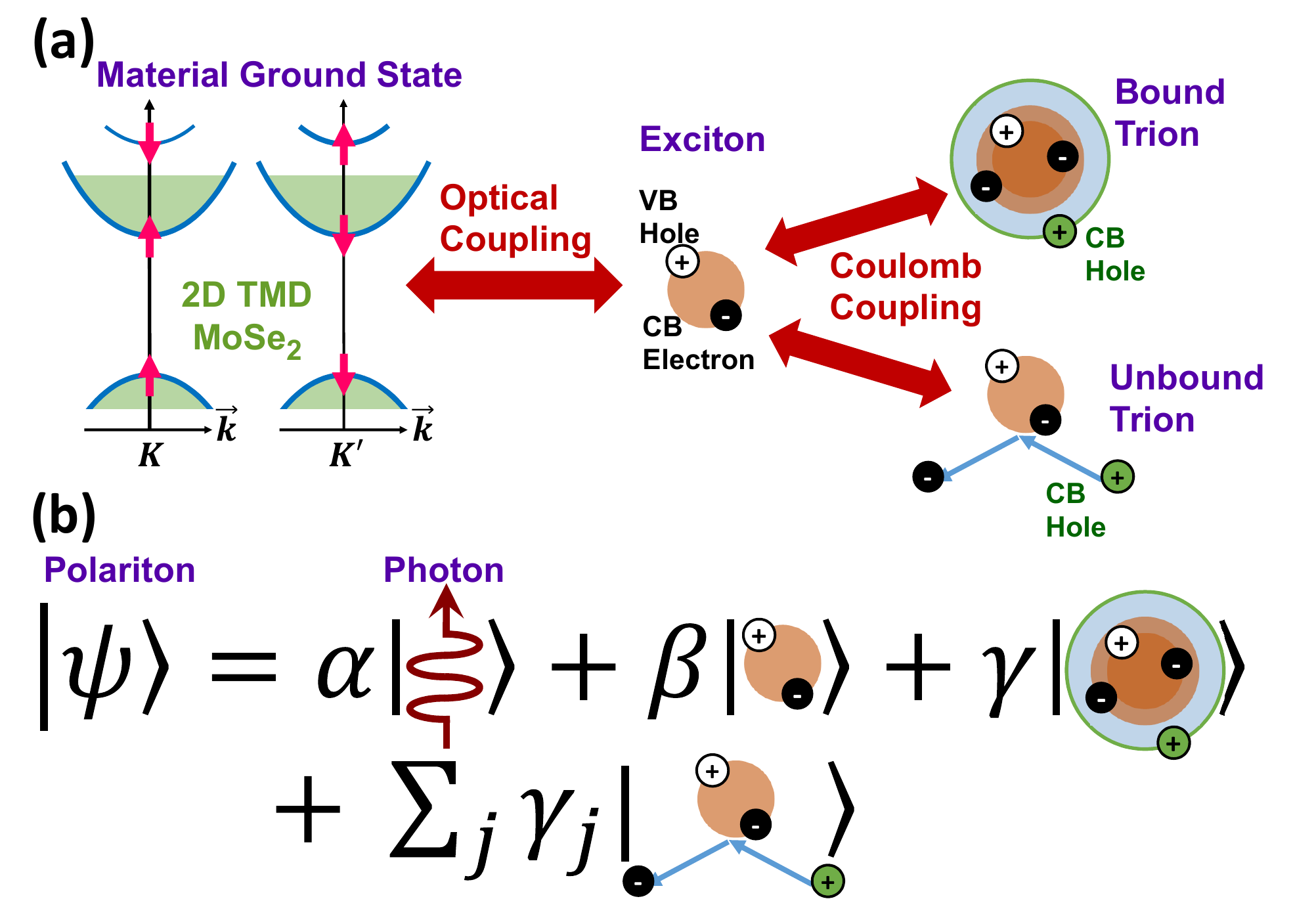}
	\caption{\label{fig:fig1} (a) The nature of couplings in involved in exciton-trion-polaritons in electron-doped 2D material MoSe\textsubscript{2} are depicted. Excitons are optically coupled to the material ground state. Excitons are also coupled to bound trions and unbound trions (exciton-electron scattering states) via Coulomb interactions. Trion states are 4-body states consisting of two conduction band (CB) electrons, one valence band (VB) hole, and one CB hole. The trion states have zero-optical matrix element with the material ground state. (b) A coherent exciton-trion-polariton state is pictorially represented as a superposition of exciton, trion, and photon states.}
\end{figure}

Recently, the authors have proposed a model in which the two peaks observed in the optical spectra of doped 2D semiconductors (the exciton and the trion, or the charged exciton, peak) are due to exciton-trion superposition states~\cite{ranamanybodytrion,ranaradlft}. Exciton states and trion states are not independent excitations in the presence of doping but are strongly coupled as a result of Coulomb interactions the strength of which depends on the doping density. Approximate energy eigenstates can be constructed using a superposition of exciton and trion states that are similar to the exciton-polaron states~\cite{atactrion,Efimkin17}. Despite the nomenclature, the trion states involved in this superposition are 4-body bosonic states and not 3-body fermionic states. The trion states have a zero optical matrix element with the material ground state and do not interact directly with light. The couplings among the exciton state, trion state, and the material ground state of an electron-doped 2D material MoSe\textsubscript{2} are depicted in Fig.~\ref{fig:fig1}(a). When such a material is placed inside a light-confining optical microstructure, the Coulomb coupling between excitons and trions and the optical coupling between excitons and photons result in coherent exciton-trion-polaritons~\cite{ranaetp}. Exciton-trion-polariton states are thus superpositions of exciton, trion, and photon states, as depicted in Fig.~\ref{fig:fig1}(b). The 4-body trion states involved in this superposition also include the continuum of exciton-electron scattering states (or unbound trion states) and their inclusion is necessary in order to get accurate results for the polariton dispersion. The 4-body bound and unbound trion states are all orthogonal to the exciton states~\cite{ranamanybodytrion}. Exciton-trion-polaritons present a novel platform to explore the many-body physics of the coupled system of excitons, trions, and photons. 

\begin{figure}[t]
	\includegraphics[width=1.0\columnwidth]{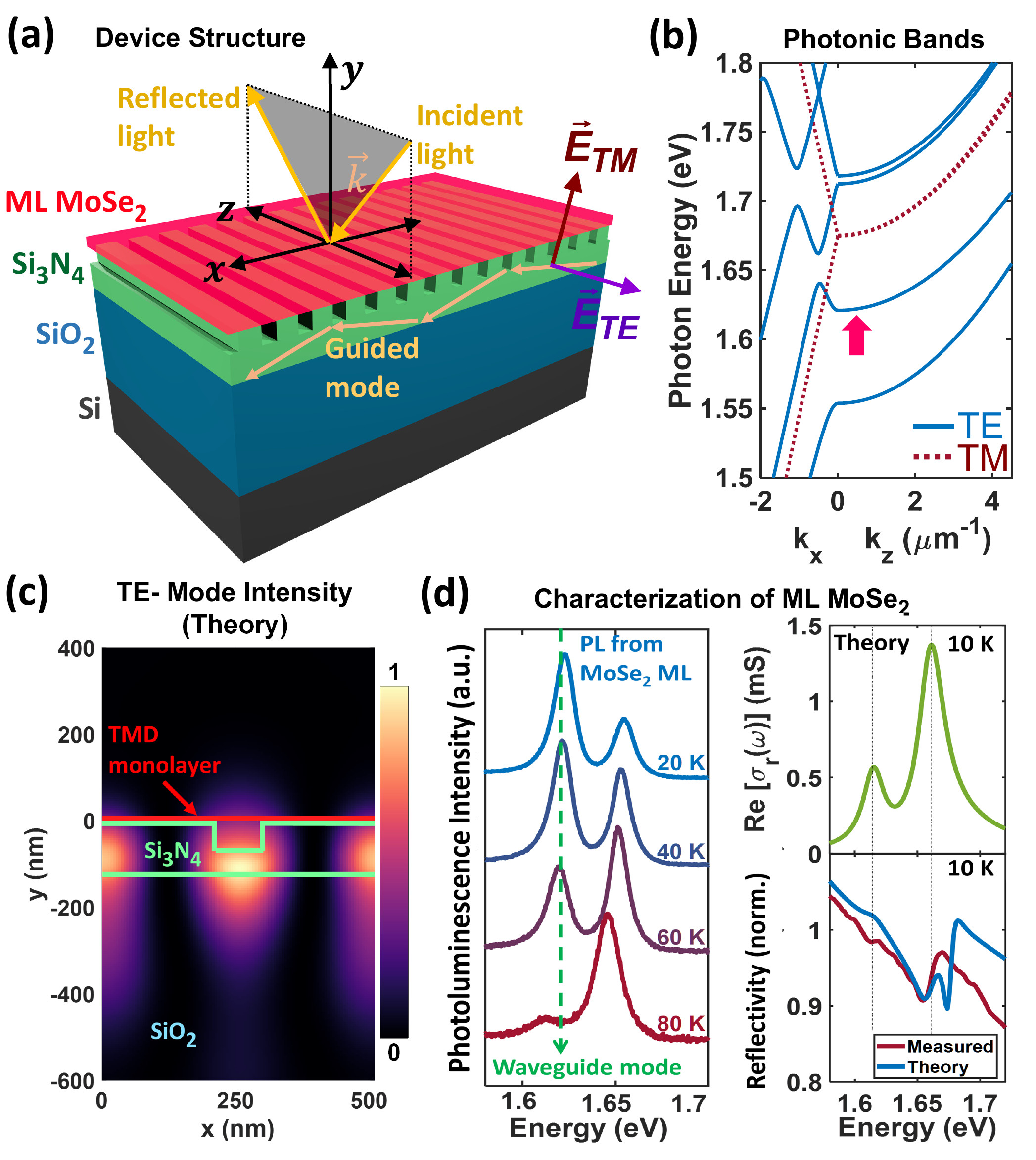}
	\caption{\label{fig:fig2} (a) Schematic of the 1D photonic crystal (PC) waveguide device used to realize exciton-trion-polaritons. The device consists of a MoSe\textsubscript{2} ML transferred on top of the waveguiding layer. (b) Band structure of the bare photonic crystal (PC) waveguide (without the MoSe\textsubscript{2} ML) is plotted as a function of the in-plane momenta. The TE-polarized waveguide mode used to realize exciton-trion-polaritons is indicated by the arrow. (c) Calculated electric field intensity of the TE-polarized waveguide mode (normalized with respect to its maximum value) is plotted for zero in-plane momentum. (d) The MoSe\textsubscript{2} ML is characterized using surface-normal photoluminescence (PL) (left plot) and reflection (bottom right plot) spectroscopies for light polarized in the x-direction. Zero-momentum energy of the TE-polarized waveguide mode is indicated by the dashed line in the PL spectra (left plot). The reflection spectra shown is normalized to the reflection spectra obtained from a part of the device which had the grating but was not covered with the MoSe\textsubscript{2} ML. The optical conductivity (real part) spectrum of the MoSe\textsubscript{2} ML, computed using the measured parameters, is also shown (top right plot).}          
\end{figure}

In this work, we experimentally realize exciton-trion-polaritons by coupling exciton-trion superposition states in an electron-doped monolayer (ML) of 2D TMD MoSe\textsubscript{2} to the optical mode in a photonic crystal (PC) waveguide (Fig.~\ref{fig:fig2}(a)) and measure the complete energy band dispersion of these polaritons. We show that our theoretical model for exciton-trion-polaritons~\cite{ranaetp}, based on the physics depicted in Fig.~\ref{fig:fig1}(a), reproduces the measured polariton energy bands and the Rabi splittings with quantitative accuracy and with no fitting parameters. The optical coupling between excitons and waveguide-confined photons, and the Coulomb coupling between excitons and trions, results in three polariton  bands; upper (UP), middle (MP), and lower (LP). Large energy splittings of $\sim$31 meV and $\sim$14 meV are observed between these three polariton bands. The measured dispersion of exciton-trion-polaritons, and the quantitative agreement between the theory and the experiments, both presented here for the first time, show that the theoretical model captures the physics associated with trion states and also with the mechanisms by which they interact with light.

\section{\label{sec:level2}Experiments, Results, and Discussion}

\begin{figure}[t]
	\includegraphics[width=0.85\columnwidth]{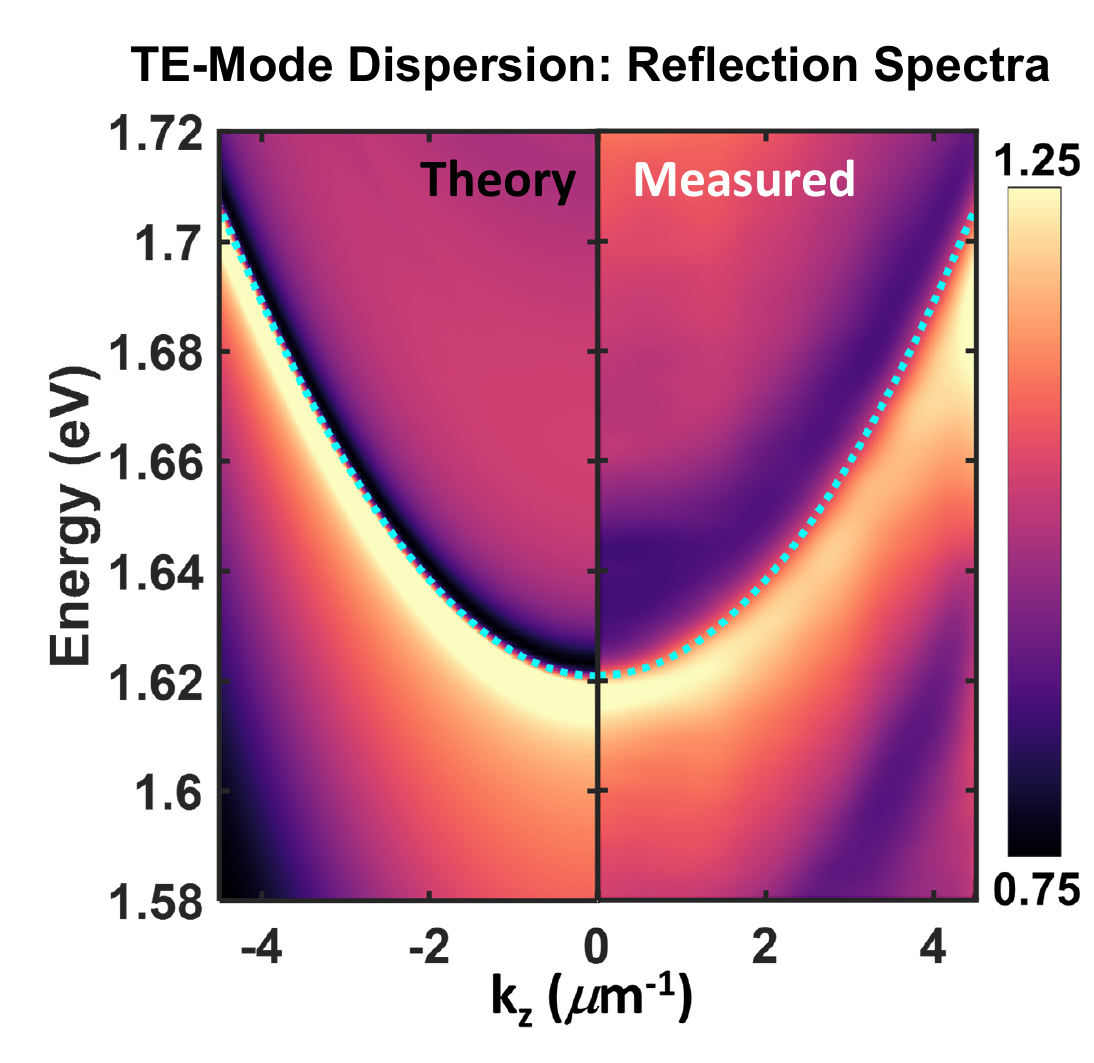}
	\caption{\label{fig:fig3} The measured (right) and calculated (left) reflection spectra of the bare photonic crystal (PC) waveguide (without the MoSe\textsubscript{2} ML) is plotted as a function of the in-plane momentum $k_{\rm z}$. The dashed lines are the computed energy-momentum dispersion of the mode. The reflection spectra shown is normalized to the reflection spectra obtained from a part of the device which had no grating and was also not covered with the MoSe\textsubscript{2} ML.}
\end{figure}

\subsection{Material and Device Characterization}
A 1D photonic crystal (PC) waveguide was realized by etching a second-order grating~\cite{Streifer89} (with a period $\Lambda=510$ nm) in a 125 nm thick Si\textsubscript{3}N\textsubscript{4} core layer deposited on top of a 1.15 $\mu$m thick SiO\textsubscript{2} cladding layer, which in turn was thermally grown on a Si substrate. A ML of TMD MoSe\textsubscript{2} was exfoliated and dry transferred on top of the Si\textsubscript{3}N\textsubscript{4} PC waveguide. The device structure is shown in Fig ~\ref{fig:fig2}(a). More details on sample preparation and device fabrication can be found in SM. Similar PC waveguide structures were used recently to explore exciton-polaritons in TMDs~\cite{dengPC}. The grating is most effective for the transverse-electric- (or TE-) polarized waveguide mode propagating in the x-direction. Since the grating is second-order, waveguide modes near wavevector $(2\pi/\Lambda)\hat{x}$ are folded into the first Brillouin zone (BZ) near the $\Gamma$-point, inside the light cone and can couple to radiation modes thereby allowing measurements of the polaritons using reflection and photoluminescence (PL) spectroscopies~\cite{Streifer89}. The calculated photonic bandstructure of the PC waveguide is shown in Fig.~\ref{fig:fig2}(b). The TE-polarized mode indicated by the arrow in Fig.~\ref{fig:fig2}(b) was used to realize the polaritons. The calculated TE-mode intensity inside the waveguide is shown in Fig.~\ref{fig:fig2}(c). Unless stated otherwise, all measurements were performed between 10 K and 20 K temperatures.


The MoSe\textsubscript{2} ML was characterized using surface-normal reflection and PL spectroscopies for light polarized in the x-direction (which does not couple to the PC waveguide modes in the wavelength range of interest) in order to determine the energies of the two peaks due to exciton-trion superposition states and their relative oscillator strengths (Fig.~\ref{fig:fig2}(d)). Since the energy splitting of the two peaks and their spectral weights are both strong functions of the doping density~\cite{ranamanybodytrion}, an electron density of $\sim 2.8 \times 10^{12}$ cm$^{-2}$ was determined from these measurements using the theoretical model for the optical conductivity of exciton-trion states developed by the authors~\cite{ranamanybodytrion}. The calculated optical conductivity (real part) for this electron density (Fig.~\ref{fig:fig2}(d)) shows that the higher energy exciton-trion spectral line has a peak oscillator strength that is $\sim$2.2 times larger than the lower energy exciton-trion spectral line. The linewidths of both the exciton-trion spectral lines were both found to be $\sim$10 meV. The TE-polarized waveguide mode, which has a calculated energy at zero wavevector equal to $\sim$1.62 eV (indicated by the dashed line in Fig.~\ref{fig:fig2}(b), left hand side plot), is nearly resonant with the lower energy exciton-trion spectral line. The energy vs in-plane momentum ($k_{\rm z}$) dispersion of this TE-polarized mode of the bare PC waveguide (without the MoSe\textsubscript{2} ML on top) was characterized using angle-dependent reflection spectroscopy. Details of the experimental setup can be found in SM. The measured momentum-dependent reflection spectrum of the mode is plotted in Fig.~\ref{fig:fig3} and shows an excellent agreement with the calculated reflection spectrum (also shown in Fig.~\ref{fig:fig3}). The dashed lines in Fig.~\ref{fig:fig3} show the mode dispersion (calculated as described in SM). Note that the mode energies are at the interface between the dark and the light regions in the reflection spectra since the  reflectivity is slightly larger (smaller) at energies below (above) the mode energies. The linewidth of the guided mode near $k_{\text{z}} = k_{\text{x}} = 0$ was determined from the reflectivity spectra to be $\sim$12 meV. This measured linewidth corresponds to a mode quality factor of $\sim$135.

\subsection{Polariton Dispersion}
The electron density in the sample and the exciton-trion linewidths, as determined from measurements described above, are the only quantities needed in our theoretical model to obtain the exciton-trion-polariton energy band dispersion, and the polariton reflection spectra, {\em without using any other fitting parameters}~\cite{ranaetp}. The computational details can be found in SM. The polariton dispersion can be obtained from the poles of the photon Green's function~\cite{ranaetp},
\begin{equation}
  G^{\rm ph}(\vec{k},\omega) = \frac{1}{\hbar \omega - \hbar \omega(\vec{k}) + i\gamma_{\rm ph} - \Sigma^{\rm ph}(\vec{k},\omega)}
\end{equation}
Here, $\vec{k}$ is the in-plane momentum, $\omega(\vec{k})$ is the optical mode frequency, $\gamma_{\rm ph}$ describes mode quality factor, and $\Sigma^{\rm ph}(\vec{k},\omega)$ is the photon self-energy that describes its interaction with the 2D ML, and equals~\cite{ranaetp},
\begin{equation}
  \Sigma^{\rm ph}(\vec{k},\omega) =\sum_{\rm s} \frac{|\hbar \Omega^{\rm ex-ph}_{\rm s}(\vec{k})/2|^{2}}{ \hbar \omega - E^{\rm ex}_{\rm s}(\vec{k}) + i\gamma_{\rm ex} - \Sigma^{\rm ex}_{\rm s}(\vec{k},\omega)}
  \end{equation}
$E^{\rm ex}_{\rm s}(\vec{k})$ is the exciton energy, $\Omega^{\rm ex-ph}_{\rm s}$ describes exciton-photon coupling~\cite{ranaetp}, $\gamma_{\rm ex}$ is the exciton decoherence rate (from all processes other than exciton-electron scattering), the subscript ${\rm s}$ stands for the spin/valley index, and $\Sigma^{\rm ph}_{\rm s}(\vec{k},\omega)$ is the exciton self-energy that describes its interaction with the trions, and equals,
\begin{equation}
  \Sigma^{\rm ex}_{\rm s}(\vec{k},\omega) = \sum_{\rm m,s'} \frac{ (1 + \delta_{\rm s,s'}) |M^{\rm ex-tr}_{\rm m,s,s'}(\vec{k})|^{2}}{ \hbar \omega - E^{\rm tr}_{\rm m,s'}(\vec{k}) + i\gamma_{\rm tr}}
\end{equation}
Here, $E^{\rm tr}_{\rm m,s'}$ is the trion energy, $M^{\rm ex-tr}_{\rm m,s,s'}$ describes the exciton-trion coupling via Coulomb interaction and is an increasing function of the doping density~\cite{ranamanybodytrion}, and $\gamma_{\rm ex}$ is the trion decoherence rate. The summation over the index ${\rm m}$ includes all bound and the continuum of unbound trion states of spin/valley index ${\rm s'}$~\cite{ranamanybodytrion}.

\begin{figure}[t]
	\includegraphics[width=0.85\columnwidth]{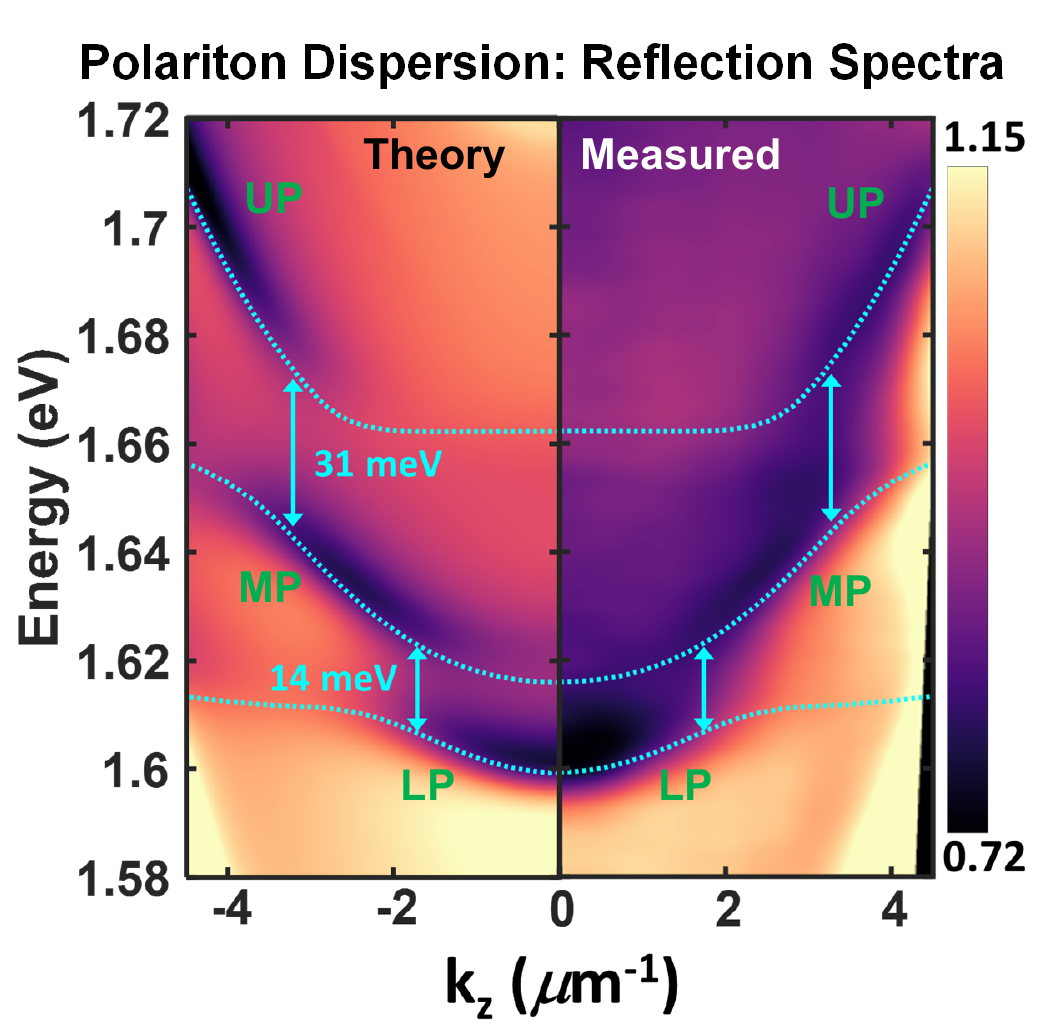}
	\caption{\label{fig:fig4} The measured (right) and calculated (left) reflection spectra of the polariton device shown in Fig.~\ref{fig:fig2}(a) is plotted as a function of the in-plane momentum $k_{\rm z}$. The dashed lines are the computed energy-momentum dispersion of the exciton-trion-polaritons. In the calculations, no parameter values were adjusted to fit the data. The reflection spectra show three polariton bands; UP, LP, and MP. The reflection spectra shown is normalized to the reflection spectra obtained from a part of the device which had no grating and was also not covered with the MoSe\textsubscript{2} ML. T=10K.} 
\end{figure}
\begin{figure}[t]
	\includegraphics[width=0.85\columnwidth]{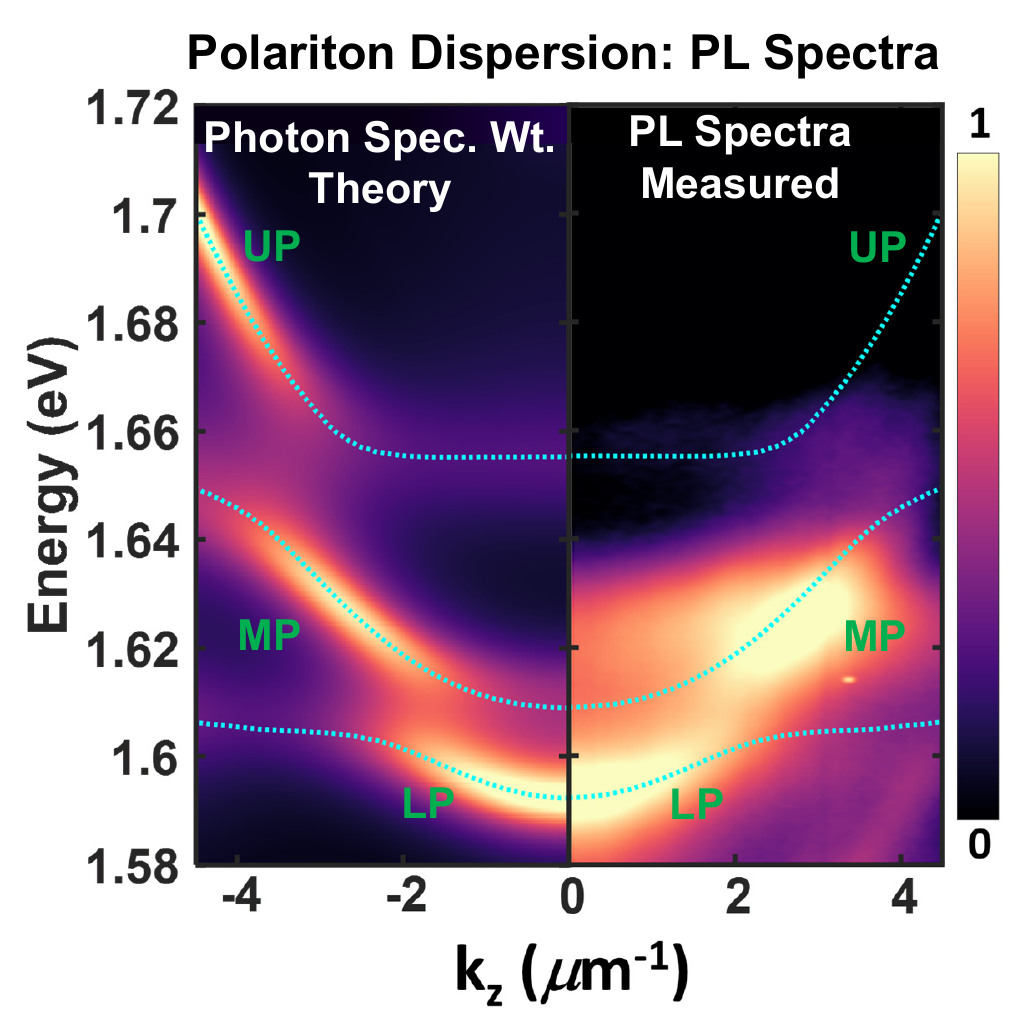}
	\caption{\label{fig:fig5} The measured PL spectra from the polariton device (right) and the computed photon spectral weight of the polariton state (left) are plotted as a function of the in-plane momentum $k_{\rm z}$. Both spectra are normalized with respect to their peak values. As expected from thermal relaxation, most of the PL emerges from the lower two polariton bands (LP and MP) and hardly any PL is obtained from the UP polariton band. T=10K} 
\end{figure}

The exciton-trion-polariton energy vs in-plane momentum ($k_{\text{z}}$) dispersion was also obtained using reflection spectroscopy. The measured reflection spectrum is plotted in Fig.~\ref{fig:fig4}. The computed reflection spectrum is also plotted in Fig.~\ref{fig:fig4} and the dashed lines show the computed polariton energy band dispersion (see SM for computational details). The measured data clearly shows three polariton bands; UP, MP, and LP in agreement with the theory~\cite{ranaetp}. The measured energy splitting between the upper two bands is $\sim$31 meV and between the lower two bands is $\sim$14 meV. These large energy splittings compared to the mode linewidth ($\sim$12 meV) and the exciton-trion linewidths ($\sim$10 meV) indicate that the polariton modes are realized in the strong coupling regime. The ratio between these two splittings is roughly equal to the ratio of the oscillator strengths of the two exciton-trion peaks in the optical conductivity spectra of the  MoSe\textsubscript{2} ML shown earlier in Fig.~\ref{fig:fig2}(d). The agreement between the theory and the experimental data in Fig.~\ref{fig:fig4} is remarkable and indicates that the theoretical model captures the essential physics associated with trions and exciton-trion-polaritons.

The reflection spectra in Fig.~\ref{fig:fig4} can be expressed in terms of the photon Green's function $G^{\text{ph}}(\vec{k},\omega)$ in the PC waveguide~\cite{GreenF76}, and the imaginary part of $G^{\text{ph}}(\vec{k},\omega)$ is related to the photon's spectral weight in the polariton~\cite{ranaetp}. More specifically, the coefficient $\alpha$ in Fig.~\ref{fig:fig1}(b) is, $|\alpha|^{2} = -\int (d \hbar\omega/\pi){\rm Im}[G^{\text{ph}}(\vec{k},\omega)]$. Consequently, the structure visible in the reflection spectra in Fig.~\ref{fig:fig4} is more pronounced where the photon contribution to the polariton is also larger. This feature is also visible in the photoluminescence (PL) spectra shown in Fig.~\ref{fig:fig5}, which plots the measured PL as a function of the in-plane momentum $k_{\text{z}}$. The strong coupling between the MoSe\textsubscript{2} ML and the waveguide optical mode results in most of the emitted PL going first into the polariton mode and then scattering out of the waveguide. The PL spectra therefore also resembles the photon's spectral weight in the polariton. The latter is also plotted in Fig.~\ref{fig:fig5}. As expected from thermal relaxation, most of the PL emerges from the lower two polariton bands (LP and MP) and hardly any PL is obtained from the UP polariton band despite pumping at intensity level $\sim$80 $\mu$W/$\mu$m$^{2}$ (with a 532 nm pump laser). At much higher pumping powers, CCD saturation effects combined with more PL collection directly from the MoSe\textsubscript{2} ML (i.e. PL which did not get coupled into the waveguide mode) result in blurring of the polariton spectral features, and much lower pumping powers result in PL coming only from the lowest energy LP band. Sample heating from the pump laser was unavoidable during PL measurements. The resulting thermal bandgap reduction of around $\sim$5 meV was incorporated into the calculations shown in Fig.~\ref{fig:fig5}. This thermal bandgap reduction is also visible in Fig.~\ref{fig:fig2}(d). Whereas all the features in the measured PL spectra of the polaritons agree well with the measured reflection spectra, the latter also provides access to the higher energy polariton bands.

\section{\label{sec:level7} Conclusion}
In this paper, we presented experimental results from exciton-trion-polaritons realized in a PC waveguide coupled to a 2D TMD MoSe\textsubscript{2} ML. The excellent agreement between the theory and the experiments provides insights into not just the structure of these polaritons but also into the nature and behavior of excitons and trions in doped semiconductors. Our work supports the argument that given the nature of the couplings (depicted in Fig.~\ref{fig:fig1}(a)), trion states by themselves cannot form polaritons (as has been assumed in in many previous works~\cite{nickTrion,Emmanuele2020,Cuadra2018,Dufferwiel2017}) and need exciton states to mediate their interaction with light and thereby contribute to polariton states. Our work also shows that given the bosonic nature of the exciton and photon states, and the experimental fact that trion states can form a coherent superposition with exciton and photon states, the trion states must themselves be bosonic. This conclusion supports the argument that the trion state in an electron-doped semiconductor is a bound state of an exciton and an electron-hole pair and is, therefore, a 4-body bound state and not a 3-body bound state, as is commonly assumed. Our work shows that polaritons can provide new windows into the physics associated with highly correlated many-body states of matter. The large binding energies of excitons and trions in 2D semiconductors, the ease with which these materials can be incorporated into optical microstructures to realize polaritons, and the charged nature of the trion states, provide fascinating opportunities to realize novel devices to exercise electrical control over electromagnetic wave propagation.

\section{\label{sec:level8} Acknowledgments}
The authors would like to acknowledge support from CCMR under NSF-NRSEC grant number DMR-1719875, support from NSF EFRI-NewLaw under grant number 1741694, and support from AFOSR  under Grant FA9550-19-1-0074.

\bibliography{references}

\providecommand{\noopsort}[1]{}\providecommand{\singleletter}[1]{#1}%
\begin{thebibliography}{34}%
\makeatletter
\providecommand \@ifxundefined [1]{%
 \@ifx{#1\undefined}
}%
\providecommand \@ifnum [1]{%
 \ifnum #1\expandafter \@firstoftwo
 \else \expandafter \@secondoftwo
 \fi
}%
\providecommand \@ifx [1]{%
 \ifx #1\expandafter \@firstoftwo
 \else \expandafter \@secondoftwo
 \fi
}%
\providecommand \natexlab [1]{#1}%
\providecommand \enquote  [1]{``#1''}%
\providecommand \bibnamefont  [1]{#1}%
\providecommand \bibfnamefont [1]{#1}%
\providecommand \citenamefont [1]{#1}%
\providecommand \href@noop [0]{\@secondoftwo}%
\providecommand \href [0]{\begingroup \@sanitize@url \@href}%
\providecommand \@href[1]{\@@startlink{#1}\@@href}%
\providecommand \@@href[1]{\endgroup#1\@@endlink}%
\providecommand \@sanitize@url [0]{\catcode `\\12\catcode `\$12\catcode
  `\&12\catcode `\#12\catcode `\^12\catcode `\_12\catcode `\%12\relax}%
\providecommand \@@startlink[1]{}%
\providecommand \@@endlink[0]{}%
\providecommand \url  [0]{\begingroup\@sanitize@url \@url }%
\providecommand \@url [1]{\endgroup\@href {#1}{\urlprefix }}%
\providecommand \urlprefix  [0]{URL }%
\providecommand \Eprint [0]{\href }%
\providecommand \doibase [0]{https://doi.org/}%
\providecommand \selectlanguage [0]{\@gobble}%
\providecommand \bibinfo  [0]{\@secondoftwo}%
\providecommand \bibfield  [0]{\@secondoftwo}%
\providecommand \translation [1]{[#1]}%
\providecommand \BibitemOpen [0]{}%
\providecommand \bibitemStop [0]{}%
\providecommand \bibitemNoStop [0]{.\EOS\space}%
\providecommand \EOS [0]{\spacefactor3000\relax}%
\providecommand \BibitemShut  [1]{\csname bibitem#1\endcsname}%
\let\auto@bib@innerbib\@empty
\bibitem [{\citenamefont {Lampert}(1958)}]{trionfirst}%
  \BibitemOpen
  \bibfield  {author} {\bibinfo {author} {\bibfnamefont {M.~A.}\ \bibnamefont
  {Lampert}},\ }\href@noop {} {\bibfield  {journal} {\bibinfo  {journal} {Phys.
  Rev. Lett.}\ }\textbf {\bibinfo {volume} {1}},\ \bibinfo {pages} {12}
  (\bibinfo {year} {1958})}\BibitemShut {NoStop}%
\bibitem [{\citenamefont {Kheng}\ \emph {et~al.}(1993)\citenamefont {Kheng},
  \citenamefont {Cox}, \citenamefont {d'~Aubign\'e}, \citenamefont {Bassani},
  \citenamefont {Saminadayar},\ and\ \citenamefont {Tatarenko}}]{trfirstexp}%
  \BibitemOpen
  \bibfield  {author} {\bibinfo {author} {\bibfnamefont {K.}~\bibnamefont
  {Kheng}}, \bibinfo {author} {\bibfnamefont {R.~T.}\ \bibnamefont {Cox}},
  \bibinfo {author} {\bibfnamefont {M.~Y.}\ \bibnamefont {d'~Aubign\'e}},
  \bibinfo {author} {\bibfnamefont {F.}~\bibnamefont {Bassani}}, \bibinfo
  {author} {\bibfnamefont {K.}~\bibnamefont {Saminadayar}},\ and\ \bibinfo
  {author} {\bibfnamefont {S.}~\bibnamefont {Tatarenko}},\ }\href
  {https://doi.org/10.1103/PhysRevLett.71.1752} {\bibfield  {journal} {\bibinfo
   {journal} {Phys. Rev. Lett.}\ }\textbf {\bibinfo {volume} {71}},\ \bibinfo
  {pages} {1752} (\bibinfo {year} {1993})}\BibitemShut {NoStop}%
\bibitem [{\citenamefont {{A. Esser, E. Runge, R. Zimmermann, W.
  Langbein}}(2000)}]{gaastrion}%
  \BibitemOpen
  \bibfield  {author} {\bibinfo {author} {\bibnamefont {{A. Esser, E. Runge, R.
  Zimmermann, W. Langbein}}},\ }\href@noop {} {\bibfield  {journal} {\bibinfo
  {journal} {Phys. Stat. Sol. (a)}\ }\textbf {\bibinfo {volume} {178}},\
  \bibinfo {pages} {489} (\bibinfo {year} {2000})}\BibitemShut {NoStop}%
\bibitem [{\citenamefont {Sergeev}\ and\ \citenamefont
  {Suris}(2001)}]{Suris01}%
  \BibitemOpen
  \bibfield  {author} {\bibinfo {author} {\bibfnamefont {R.~A.}\ \bibnamefont
  {Sergeev}}\ and\ \bibinfo {author} {\bibfnamefont {R.~A.}\ \bibnamefont
  {Suris}},\ }\href {https://doi.org/10.1134/1.1366005} {\bibfield  {journal}
  {\bibinfo  {journal} {Phys. Solid State}\ }\textbf {\bibinfo {volume} {43}},\
  \bibinfo {pages} {746} (\bibinfo {year} {2001})}\BibitemShut {NoStop}%
\bibitem [{\citenamefont {Combescot}\ and\ \citenamefont
  {Tribollet}(2003)}]{Ctrion0}%
  \BibitemOpen
  \bibfield  {author} {\bibinfo {author} {\bibfnamefont {M.}~\bibnamefont
  {Combescot}}\ and\ \bibinfo {author} {\bibfnamefont {J.}~\bibnamefont
  {Tribollet}},\ }\href
  {https://doi.org/https://doi.org/10.1016/S0038-1098(03)00657-4} {\bibfield
  {journal} {\bibinfo  {journal} {Solid State Commun.}\ }\textbf {\bibinfo
  {volume} {128}},\ \bibinfo {pages} {273} (\bibinfo {year}
  {2003})}\BibitemShut {NoStop}%
\bibitem [{\citenamefont {Berkelbach}\ \emph {et~al.}(2013)\citenamefont
  {Berkelbach}, \citenamefont {Hybertsen},\ and\ \citenamefont
  {Reichman}}]{Berk13}%
  \BibitemOpen
  \bibfield  {author} {\bibinfo {author} {\bibfnamefont {T.~C.}\ \bibnamefont
  {Berkelbach}}, \bibinfo {author} {\bibfnamefont {M.~S.}\ \bibnamefont
  {Hybertsen}},\ and\ \bibinfo {author} {\bibfnamefont {D.~R.}\ \bibnamefont
  {Reichman}},\ }\href {https://doi.org/10.1103/PhysRevB.88.045318} {\bibfield
  {journal} {\bibinfo  {journal} {Phys. Rev. B}\ }\textbf {\bibinfo {volume}
  {88}},\ \bibinfo {pages} {045318} (\bibinfo {year} {2013})}\BibitemShut
  {NoStop}%
\bibitem [{\citenamefont {Zhang}\ \emph {et~al.}(2014)\citenamefont {Zhang},
  \citenamefont {Wang}, \citenamefont {Chan}, \citenamefont {Manolatou},\ and\
  \citenamefont {Rana}}]{Rana14}%
  \BibitemOpen
  \bibfield  {author} {\bibinfo {author} {\bibfnamefont {C.}~\bibnamefont
  {Zhang}}, \bibinfo {author} {\bibfnamefont {H.}~\bibnamefont {Wang}},
  \bibinfo {author} {\bibfnamefont {W.}~\bibnamefont {Chan}}, \bibinfo {author}
  {\bibfnamefont {C.}~\bibnamefont {Manolatou}},\ and\ \bibinfo {author}
  {\bibfnamefont {F.}~\bibnamefont {Rana}},\ }\href
  {https://doi.org/10.1103/PhysRevB.89.205436} {\bibfield  {journal} {\bibinfo
  {journal} {Phys. Rev. B}\ }\textbf {\bibinfo {volume} {89}},\ \bibinfo
  {pages} {205436} (\bibinfo {year} {2014})}\BibitemShut {NoStop}%
\bibitem [{\citenamefont {Courtade}\ \emph {et~al.}(2017)\citenamefont
  {Courtade}, \citenamefont {Semina}, \citenamefont {Manca}, \citenamefont
  {Glazov}, \citenamefont {Robert}, \citenamefont {Cadiz}, \citenamefont
  {Wang}, \citenamefont {Taniguchi}, \citenamefont {Watanabe}, \citenamefont
  {Pierre}, \citenamefont {Escoffier}, \citenamefont {Ivchenko}, \citenamefont
  {Renucci}, \citenamefont {Marie}, \citenamefont {Amand},\ and\ \citenamefont
  {Urbaszek}}]{Urba17}%
  \BibitemOpen
  \bibfield  {author} {\bibinfo {author} {\bibfnamefont {E.}~\bibnamefont
  {Courtade}}, \bibinfo {author} {\bibfnamefont {M.}~\bibnamefont {Semina}},
  \bibinfo {author} {\bibfnamefont {M.}~\bibnamefont {Manca}}, \bibinfo
  {author} {\bibfnamefont {M.~M.}\ \bibnamefont {Glazov}}, \bibinfo {author}
  {\bibfnamefont {C.}~\bibnamefont {Robert}}, \bibinfo {author} {\bibfnamefont
  {F.}~\bibnamefont {Cadiz}}, \bibinfo {author} {\bibfnamefont
  {G.}~\bibnamefont {Wang}}, \bibinfo {author} {\bibfnamefont {T.}~\bibnamefont
  {Taniguchi}}, \bibinfo {author} {\bibfnamefont {K.}~\bibnamefont {Watanabe}},
  \bibinfo {author} {\bibfnamefont {M.}~\bibnamefont {Pierre}}, \bibinfo
  {author} {\bibfnamefont {W.}~\bibnamefont {Escoffier}}, \bibinfo {author}
  {\bibfnamefont {E.~L.}\ \bibnamefont {Ivchenko}}, \bibinfo {author}
  {\bibfnamefont {P.}~\bibnamefont {Renucci}}, \bibinfo {author} {\bibfnamefont
  {X.}~\bibnamefont {Marie}}, \bibinfo {author} {\bibfnamefont
  {T.}~\bibnamefont {Amand}},\ and\ \bibinfo {author} {\bibfnamefont
  {B.}~\bibnamefont {Urbaszek}},\ }\href
  {https://doi.org/10.1103/PhysRevB.96.085302} {\bibfield  {journal} {\bibinfo
  {journal} {Phys. Rev. B}\ }\textbf {\bibinfo {volume} {96}},\ \bibinfo
  {pages} {085302} (\bibinfo {year} {2017})}\BibitemShut {NoStop}%
\bibitem [{\citenamefont {Esser}\ \emph {et~al.}(2001)\citenamefont {Esser},
  \citenamefont {Zimmermann},\ and\ \citenamefont {Runge}}]{EsserTrion}%
  \BibitemOpen
  \bibfield  {author} {\bibinfo {author} {\bibfnamefont {A.}~\bibnamefont
  {Esser}}, \bibinfo {author} {\bibfnamefont {R.}~\bibnamefont {Zimmermann}},\
  and\ \bibinfo {author} {\bibfnamefont {E.}~\bibnamefont {Runge}},\
  }\href@noop {} {\bibfield  {journal} {\bibinfo  {journal} {Phys. Status
  Solidi B}\ }\textbf {\bibinfo {volume} {227}},\ \bibinfo {pages} {317}
  (\bibinfo {year} {2001})}\BibitemShut {NoStop}%
\bibitem [{\citenamefont {{M. Combescot, O. Betbeder-Matibet, F.
  Dubin}}(2004)}]{Ctrion}%
  \BibitemOpen
  \bibfield  {author} {\bibinfo {author} {\bibnamefont {{M. Combescot, O.
  Betbeder-Matibet, F. Dubin}}},\ }\href@noop {} {\bibfield  {journal}
  {\bibinfo  {journal} {Eur. Phys. J. B}\ }\textbf {\bibinfo {volume} {42}},\
  \bibinfo {pages} {63} (\bibinfo {year} {2004})}\BibitemShut {NoStop}%
\bibitem [{\citenamefont {{Shiue-Yuan Shiau, Monique Combescot, Yia-Chung
  Chang}}(2012)}]{Crecent}%
  \BibitemOpen
  \bibfield  {author} {\bibinfo {author} {\bibnamefont {{Shiue-Yuan Shiau,
  Monique Combescot, Yia-Chung Chang}}},\ }\href@noop {} {\bibfield  {journal}
  {\bibinfo  {journal} {Phys. Rev. B}\ }\textbf {\bibinfo {volume} {86}},\
  \bibinfo {pages} {115210} (\bibinfo {year} {2012})}\BibitemShut {NoStop}%
\bibitem [{\citenamefont {Combescot}\ and\ \citenamefont
  {Betbeder-Matibet}(2003)}]{Ctrion2}%
  \BibitemOpen
  \bibfield  {author} {\bibinfo {author} {\bibfnamefont {M.}~\bibnamefont
  {Combescot}}\ and\ \bibinfo {author} {\bibfnamefont {O.}~\bibnamefont
  {Betbeder-Matibet}},\ }\href@noop {} {\bibfield  {journal} {\bibinfo
  {journal} {Solid State Commun.}\ }\textbf {\bibinfo {volume} {126}},\
  \bibinfo {pages} {687 } (\bibinfo {year} {2003})}\BibitemShut {NoStop}%
\bibitem [{\citenamefont {Kyriienko}\ \emph {et~al.}(2020)\citenamefont
  {Kyriienko}, \citenamefont {Krizhanovskii},\ and\ \citenamefont
  {Shelykh}}]{Nonlineartr20}%
  \BibitemOpen
  \bibfield  {author} {\bibinfo {author} {\bibfnamefont {O.}~\bibnamefont
  {Kyriienko}}, \bibinfo {author} {\bibfnamefont {D.~N.}\ \bibnamefont
  {Krizhanovskii}},\ and\ \bibinfo {author} {\bibfnamefont {I.~A.}\
  \bibnamefont {Shelykh}},\ }\href
  {https://doi.org/10.1103/PhysRevLett.125.197402} {\bibfield  {journal}
  {\bibinfo  {journal} {Phys. Rev. Lett.}\ }\textbf {\bibinfo {volume} {125}},\
  \bibinfo {pages} {197402} (\bibinfo {year} {2020})}\BibitemShut {NoStop}%
\bibitem [{\citenamefont {{S. Dhara, C. Chakraborty, K. M. Goodfellow, L. Qiu,
  T. A. O'Loughlin, G. W. Wicks, Subhro Bhattacharjee, A. N.
  Vamivakas}}(2017)}]{nickTrion}%
  \BibitemOpen
  \bibfield  {author} {\bibinfo {author} {\bibnamefont {{S. Dhara, C.
  Chakraborty, K. M. Goodfellow, L. Qiu, T. A. O'Loughlin, G. W. Wicks, Subhro
  Bhattacharjee, A. N. Vamivakas}}},\ }\href@noop {} {\bibfield  {journal}
  {\bibinfo  {journal} {Nat. Phys.}\ }\textbf {\bibinfo {volume} {14}},\
  \bibinfo {pages} {130} (\bibinfo {year} {2017})}\BibitemShut {NoStop}%
\bibitem [{\citenamefont {Emmanuele}\ \emph {et~al.}(2020)\citenamefont
  {Emmanuele}, \citenamefont {Sich}, \citenamefont {Kyriienko}, \citenamefont
  {Shahnazaryan}, \citenamefont {Withers}, \citenamefont {Catanzaro},
  \citenamefont {Walker}, \citenamefont {Benimetskiy}, \citenamefont
  {Skolnick}, \citenamefont {Tartakovskii}, \citenamefont {Shelykh},\ and\
  \citenamefont {Krizhanovskii}}]{Emmanuele2020}%
  \BibitemOpen
  \bibfield  {author} {\bibinfo {author} {\bibfnamefont {R.~P.~A.}\
  \bibnamefont {Emmanuele}}, \bibinfo {author} {\bibfnamefont {M.}~\bibnamefont
  {Sich}}, \bibinfo {author} {\bibfnamefont {O.}~\bibnamefont {Kyriienko}},
  \bibinfo {author} {\bibfnamefont {V.}~\bibnamefont {Shahnazaryan}}, \bibinfo
  {author} {\bibfnamefont {F.}~\bibnamefont {Withers}}, \bibinfo {author}
  {\bibfnamefont {A.}~\bibnamefont {Catanzaro}}, \bibinfo {author}
  {\bibfnamefont {P.~M.}\ \bibnamefont {Walker}}, \bibinfo {author}
  {\bibfnamefont {F.~A.}\ \bibnamefont {Benimetskiy}}, \bibinfo {author}
  {\bibfnamefont {M.~S.}\ \bibnamefont {Skolnick}}, \bibinfo {author}
  {\bibfnamefont {A.~I.}\ \bibnamefont {Tartakovskii}}, \bibinfo {author}
  {\bibfnamefont {I.~A.}\ \bibnamefont {Shelykh}},\ and\ \bibinfo {author}
  {\bibfnamefont {D.~N.}\ \bibnamefont {Krizhanovskii}},\ }\href
  {https://doi.org/10.1038/s41467-020-17340-z} {\bibfield  {journal} {\bibinfo
  {journal} {Nat. Commun}\ }\textbf {\bibinfo {volume} {11}},\ \bibinfo {pages}
  {3589} (\bibinfo {year} {2020})}\BibitemShut {NoStop}%
\bibitem [{\citenamefont {Cuadra}\ \emph {et~al.}(2018)\citenamefont {Cuadra},
  \citenamefont {Baranov}, \citenamefont {Wers{\"a}ll}, \citenamefont {Verre},
  \citenamefont {Antosiewicz},\ and\ \citenamefont {Shegai}}]{Cuadra2018}%
  \BibitemOpen
  \bibfield  {author} {\bibinfo {author} {\bibfnamefont {J.}~\bibnamefont
  {Cuadra}}, \bibinfo {author} {\bibfnamefont {D.~G.}\ \bibnamefont {Baranov}},
  \bibinfo {author} {\bibfnamefont {M.}~\bibnamefont {Wers{\"a}ll}}, \bibinfo
  {author} {\bibfnamefont {R.}~\bibnamefont {Verre}}, \bibinfo {author}
  {\bibfnamefont {T.~J.}\ \bibnamefont {Antosiewicz}},\ and\ \bibinfo {author}
  {\bibfnamefont {T.}~\bibnamefont {Shegai}},\ }\href
  {https://doi.org/10.1021/acs.nanolett.7b04965} {\bibfield  {journal}
  {\bibinfo  {journal} {Nano Lett.}\ }\textbf {\bibinfo {volume} {18}},\
  \bibinfo {pages} {1777} (\bibinfo {year} {2018})}\BibitemShut {NoStop}%
\bibitem [{\citenamefont {Dufferwiel}\ \emph {et~al.}(2017)\citenamefont
  {Dufferwiel}, \citenamefont {Lyons}, \citenamefont {Solnyshkov},
  \citenamefont {Trichet}, \citenamefont {Withers}, \citenamefont {Schwarz},
  \citenamefont {Malpuech}, \citenamefont {Smith}, \citenamefont {Novoselov},
  \citenamefont {Skolnick}, \citenamefont {Krizhanovskii},\ and\ \citenamefont
  {Tartakovskii}}]{Dufferwiel2017}%
  \BibitemOpen
  \bibfield  {author} {\bibinfo {author} {\bibfnamefont {S.}~\bibnamefont
  {Dufferwiel}}, \bibinfo {author} {\bibfnamefont {T.~P.}\ \bibnamefont
  {Lyons}}, \bibinfo {author} {\bibfnamefont {D.~D.}\ \bibnamefont
  {Solnyshkov}}, \bibinfo {author} {\bibfnamefont {A.~A.~P.}\ \bibnamefont
  {Trichet}}, \bibinfo {author} {\bibfnamefont {F.}~\bibnamefont {Withers}},
  \bibinfo {author} {\bibfnamefont {S.}~\bibnamefont {Schwarz}}, \bibinfo
  {author} {\bibfnamefont {G.}~\bibnamefont {Malpuech}}, \bibinfo {author}
  {\bibfnamefont {J.~M.}\ \bibnamefont {Smith}}, \bibinfo {author}
  {\bibfnamefont {K.~S.}\ \bibnamefont {Novoselov}}, \bibinfo {author}
  {\bibfnamefont {M.~S.}\ \bibnamefont {Skolnick}}, \bibinfo {author}
  {\bibfnamefont {D.~N.}\ \bibnamefont {Krizhanovskii}},\ and\ \bibinfo
  {author} {\bibfnamefont {A.~I.}\ \bibnamefont {Tartakovskii}},\ }\href
  {https://doi.org/10.1038/nphoton.2017.125} {\bibfield  {journal} {\bibinfo
  {journal} {Nat. Photonics}\ }\textbf {\bibinfo {volume} {11}},\ \bibinfo
  {pages} {497} (\bibinfo {year} {2017})}\BibitemShut {NoStop}%
\bibitem [{\citenamefont {{Farhan Rana, Okan Koksal, Christina
  Manolatou}}(2020)}]{ranamanybodytrion}%
  \BibitemOpen
  \bibfield  {author} {\bibinfo {author} {\bibnamefont {{Farhan Rana, Okan
  Koksal, Christina Manolatou}}},\ }\href@noop {} {\bibfield  {journal}
  {\bibinfo  {journal} {Phys. Rev. B}\ }\textbf {\bibinfo {volume} {102}},\
  \bibinfo {pages} {085304} (\bibinfo {year} {2020})}\BibitemShut {NoStop}%
\bibitem [{\citenamefont {Rana}\ \emph
  {et~al.}(2021{\natexlab{a}})\citenamefont {Rana}, \citenamefont {Koksal},
  \citenamefont {Jung}, \citenamefont {Shvets},\ and\ \citenamefont
  {Manolatou}}]{ranaradlft}%
  \BibitemOpen
  \bibfield  {author} {\bibinfo {author} {\bibfnamefont {F.}~\bibnamefont
  {Rana}}, \bibinfo {author} {\bibfnamefont {O.}~\bibnamefont {Koksal}},
  \bibinfo {author} {\bibfnamefont {M.}~\bibnamefont {Jung}}, \bibinfo {author}
  {\bibfnamefont {G.}~\bibnamefont {Shvets}},\ and\ \bibinfo {author}
  {\bibfnamefont {C.}~\bibnamefont {Manolatou}},\ }\href
  {https://doi.org/10.1103/PhysRevB.103.035424} {\bibfield  {journal} {\bibinfo
   {journal} {Phys. Rev. B}\ }\textbf {\bibinfo {volume} {103}},\ \bibinfo
  {pages} {035424} (\bibinfo {year} {2021}{\natexlab{a}})}\BibitemShut
  {NoStop}%
\bibitem [{\citenamefont {{Meinrad Sidler, Patrick Back, Ovidiu Cotlet, Ajit
  Sristava, Thomas Fink, Martin Kroner, Eugene Demler, Atac
  Imamoglu}}(2016)}]{atactrion}%
  \BibitemOpen
  \bibfield  {author} {\bibinfo {author} {\bibnamefont {{Meinrad Sidler,
  Patrick Back, Ovidiu Cotlet, Ajit Sristava, Thomas Fink, Martin Kroner,
  Eugene Demler, Atac Imamoglu}}},\ }\href@noop {} {\bibfield  {journal}
  {\bibinfo  {journal} {Nat. Phys}\ }\textbf {\bibinfo {volume} {13}},\
  \bibinfo {pages} {255} (\bibinfo {year} {2016})}\BibitemShut {NoStop}%
\bibitem [{\citenamefont {{D. K. Efimkin, A. H. MacDonald}}(2017)}]{Efimkin17}%
  \BibitemOpen
  \bibfield  {author} {\bibinfo {author} {\bibnamefont {{D. K. Efimkin, A. H.
  MacDonald}}},\ }\href@noop {} {\bibfield  {journal} {\bibinfo  {journal}
  {Phys. Rev. B}\ }\textbf {\bibinfo {volume} {95}},\ \bibinfo {pages} {035417}
  (\bibinfo {year} {2017})}\BibitemShut {NoStop}%
\bibitem [{\citenamefont {Rana}\ \emph
  {et~al.}(2021{\natexlab{b}})\citenamefont {Rana}, \citenamefont {Koksal},
  \citenamefont {Jung}, \citenamefont {Shvets}, \citenamefont {Vamivakas},\
  and\ \citenamefont {Manolatou}}]{ranaetp}%
  \BibitemOpen
  \bibfield  {author} {\bibinfo {author} {\bibfnamefont {F.}~\bibnamefont
  {Rana}}, \bibinfo {author} {\bibfnamefont {O.}~\bibnamefont {Koksal}},
  \bibinfo {author} {\bibfnamefont {M.}~\bibnamefont {Jung}}, \bibinfo {author}
  {\bibfnamefont {G.}~\bibnamefont {Shvets}}, \bibinfo {author} {\bibfnamefont
  {A.~N.}\ \bibnamefont {Vamivakas}},\ and\ \bibinfo {author} {\bibfnamefont
  {C.}~\bibnamefont {Manolatou}},\ }\href
  {https://doi.org/10.1103/PhysRevLett.126.127402} {\bibfield  {journal}
  {\bibinfo  {journal} {Phys. Rev. Lett.}\ }\textbf {\bibinfo {volume} {126}},\
  \bibinfo {pages} {127402} (\bibinfo {year} {2021}{\natexlab{b}})}\BibitemShut
  {NoStop}%
\bibitem [{\citenamefont {{Hardy}}\ \emph {et~al.}(1989)\citenamefont
  {{Hardy}}, \citenamefont {{Welch}},\ and\ \citenamefont
  {{Streifer}}}]{Streifer89}%
  \BibitemOpen
  \bibfield  {author} {\bibinfo {author} {\bibfnamefont {A.}~\bibnamefont
  {{Hardy}}}, \bibinfo {author} {\bibfnamefont {D.~F.}\ \bibnamefont
  {{Welch}}},\ and\ \bibinfo {author} {\bibfnamefont {W.}~\bibnamefont
  {{Streifer}}},\ }\href {https://doi.org/10.1109/3.35721} {\bibfield
  {journal} {\bibinfo  {journal} {IEEE J. Quantum Electron.}\ }\textbf
  {\bibinfo {volume} {25}},\ \bibinfo {pages} {2096} (\bibinfo {year}
  {1989})}\BibitemShut {NoStop}%
\bibitem [{\citenamefont {{Long Zhang, Rahul Gogna, Will Burg, Emanuel Tutuc,
  Hui Deng}}(2018)}]{dengPC}%
  \BibitemOpen
  \bibfield  {author} {\bibinfo {author} {\bibnamefont {{Long Zhang, Rahul
  Gogna, Will Burg, Emanuel Tutuc, Hui Deng}}},\ }\href@noop {} {\bibfield
  {journal} {\bibinfo  {journal} {Nat. Commun}\ }\textbf {\bibinfo {volume}
  {9}},\ \bibinfo {pages} {713} (\bibinfo {year} {2018})}\BibitemShut {NoStop}%
\bibitem [{\citenamefont {{Ghizoni}}\ \emph {et~al.}(1976)\citenamefont
  {{Ghizoni}}, \citenamefont {{Bor-Uei Chen}},\ and\ \citenamefont {{Chung
  Tang}}}]{GreenF76}%
  \BibitemOpen
  \bibfield  {author} {\bibinfo {author} {\bibfnamefont {C.}~\bibnamefont
  {{Ghizoni}}}, \bibinfo {author} {\bibnamefont {{Bor-Uei Chen}}},\ and\
  \bibinfo {author} {\bibnamefont {{Chung Tang}}},\ }\href
  {https://doi.org/10.1109/JQE.1976.1069099} {\bibfield  {journal} {\bibinfo
  {journal} {IEEE J. Quantum Electron.}\ }\textbf {\bibinfo {volume} {12}},\
  \bibinfo {pages} {69} (\bibinfo {year} {1976})}\BibitemShut {NoStop}%
\bibitem [{\citenamefont {{Yong-Jiu Zhao}}\ \emph {et~al.}(2002)\citenamefont
  {{Yong-Jiu Zhao}}, \citenamefont {{Ke-Li Wu}},\ and\ \citenamefont
  {{Kwok-Keung M. Cheng}}}]{FDFD}%
  \BibitemOpen
  \bibfield  {author} {\bibinfo {author} {\bibnamefont {{Yong-Jiu Zhao}}},
  \bibinfo {author} {\bibnamefont {{Ke-Li Wu}}},\ and\ \bibinfo {author}
  {\bibnamefont {{Kwok-Keung M. Cheng}}},\ }\href
  {https://doi.org/10.1109/TMTT.2002.800447} {\bibfield  {journal} {\bibinfo
  {journal} {IEEE Trans Microw Theory Tech}\ }\textbf {\bibinfo {volume}
  {50}},\ \bibinfo {pages} {1844} (\bibinfo {year} {2002})}\BibitemShut
  {NoStop}%
\bibitem [{\citenamefont {{Fabian Cadiz, Cedric Robert, Gang Wang, Wilson Kong,
  Xi Fan, Mark Blei, Delphine Lagarde, Maxime Gay, Takahashi Taniguchi, Kenji
  Watanabe, Thierry Amand, Xavier Marie, Pierre Renucci, Sefaattin Tongay,
  Bernhard Urbaszek}}(2016)}]{mose2thickness}%
  \BibitemOpen
  \bibfield  {author} {\bibinfo {author} {\bibnamefont {{Fabian Cadiz, Cedric
  Robert, Gang Wang, Wilson Kong, Xi Fan, Mark Blei, Delphine Lagarde, Maxime
  Gay, Takahashi Taniguchi, Kenji Watanabe, Thierry Amand, Xavier Marie, Pierre
  Renucci, Sefaattin Tongay, Bernhard Urbaszek}}},\ }\href@noop {} {\bibfield
  {journal} {\bibinfo  {journal} {2D Mater.}\ }\textbf {\bibinfo {volume}
  {3}},\ \bibinfo {pages} {045008} (\bibinfo {year} {2016})}\BibitemShut
  {NoStop}%
\bibitem [{\citenamefont {{Yifei Li, Alexey Chernikov, Xian Zhang, Albert
  Rigosi, Heather M. Hill, Arend M. van der Zande, Daniel A. Chenet, En-Min
  Shih, James Hone, Tony F. Heinz}}(2014)}]{mose2permittivity}%
  \BibitemOpen
  \bibfield  {author} {\bibinfo {author} {\bibnamefont {{Yifei Li, Alexey
  Chernikov, Xian Zhang, Albert Rigosi, Heather M. Hill, Arend M. van der
  Zande, Daniel A. Chenet, En-Min Shih, James Hone, Tony F. Heinz}}},\
  }\href@noop {} {\bibfield  {journal} {\bibinfo  {journal} {Phys. Rev. B}\
  }\textbf {\bibinfo {volume} {90}},\ \bibinfo {pages} {205422} (\bibinfo
  {year} {2014})}\BibitemShut {NoStop}%
\bibitem [{\citenamefont {Monique~Combescot}(2015)}]{Ctrion3}%
  \BibitemOpen
  \bibfield  {author} {\bibinfo {author} {\bibfnamefont {S.-Y.~S.}\
  \bibnamefont {Monique~Combescot}},\ }\href@noop {} {\emph {\bibinfo {title}
  {Excitons and Cooper Pairs: Two Composite Bosons in Many-Body Physics}}}\
  (\bibinfo  {publisher} {Oxford Graduate Texts},\ \bibinfo {year}
  {2015})\BibitemShut {NoStop}%
\bibitem [{\citenamefont {Glazov}(2020)}]{Glazov}%
  \BibitemOpen
  \bibfield  {author} {\bibinfo {author} {\bibfnamefont {M.~M.}\ \bibnamefont
  {Glazov}},\ }\href {https://doi.org/10.1063/5.0012475} {\bibfield  {journal}
  {\bibinfo  {journal} {J. Chem. Phys.}\ }\textbf {\bibinfo {volume} {153}},\
  \bibinfo {pages} {034703} (\bibinfo {year} {2020})}\BibitemShut {NoStop}%
\bibitem [{\citenamefont {Rapaport}\ \emph {et~al.}(2001)\citenamefont
  {Rapaport}, \citenamefont {Cohen}, \citenamefont {Ron}, \citenamefont
  {Linder},\ and\ \citenamefont {Pfeiffer}}]{SPfeiffer01}%
  \BibitemOpen
  \bibfield  {author} {\bibinfo {author} {\bibfnamefont {R.}~\bibnamefont
  {Rapaport}}, \bibinfo {author} {\bibfnamefont {E.}~\bibnamefont {Cohen}},
  \bibinfo {author} {\bibfnamefont {A.}~\bibnamefont {Ron}}, \bibinfo {author}
  {\bibfnamefont {E.}~\bibnamefont {Linder}},\ and\ \bibinfo {author}
  {\bibfnamefont {L.~N.}\ \bibnamefont {Pfeiffer}},\ }\href
  {https://doi.org/10.1103/PhysRevB.63.235310} {\bibfield  {journal} {\bibinfo
  {journal} {Phys. Rev. B}\ }\textbf {\bibinfo {volume} {63}} (\bibinfo {year}
  {2001})}\BibitemShut {NoStop}%
\bibitem [{\citenamefont {{G. Plechinger, P. Nagler, A. Arora, R. Schmidt, A.
  Chernikov, A. G. del Aguila, P. C. M. Christianen, R. Bratschitsch, C.
  Schuller, T. Korn}}(2016)}]{Korn16}%
  \BibitemOpen
  \bibfield  {author} {\bibinfo {author} {\bibnamefont {{G. Plechinger, P.
  Nagler, A. Arora, R. Schmidt, A. Chernikov, A. G. del Aguila, P. C. M.
  Christianen, R. Bratschitsch, C. Schuller, T. Korn}}},\ }\href@noop {}
  {\bibfield  {journal} {\bibinfo  {journal} {Nat. Commun}\ }\textbf {\bibinfo
  {volume} {7}},\ \bibinfo {pages} {12715} (\bibinfo {year}
  {2016})}\BibitemShut {NoStop}%
\bibitem [{\citenamefont {{Richard Schmidt, Michael Knap, Dmitri Ivanov,
  Jhih-Shih You, Marko Cetina, Eugene Demler}}(2018)}]{demreview}%
  \BibitemOpen
  \bibfield  {author} {\bibinfo {author} {\bibnamefont {{Richard Schmidt,
  Michael Knap, Dmitri Ivanov, Jhih-Shih You, Marko Cetina, Eugene Demler}}},\
  }\href@noop {} {\bibfield  {journal} {\bibinfo  {journal} {Rep. Prog. Phys.}\
  }\textbf {\bibinfo {volume} {81}},\ \bibinfo {pages} {024401} (\bibinfo
  {year} {2018})}\BibitemShut {NoStop}%
\bibitem [{\citenamefont {{F. Chevy}}(2006)}]{Chevy06}%
  \BibitemOpen
  \bibfield  {author} {\bibinfo {author} {\bibnamefont {{F. Chevy}}},\
  }\href@noop {} {\bibfield  {journal} {\bibinfo  {journal} {Phys. Rev. A}\
  }\textbf {\bibinfo {volume} {74}},\ \bibinfo {pages} {063628} (\bibinfo
  {year} {2006})}\BibitemShut {NoStop}%
\end{thebibliography}%

\section*{\label{sec:Supp1} Supplementary Material}

\bibliographystyle{apsrev4-2}

\subsection*{\label{sec:level10} Sample Fabrication Details}
The device structure shown in Fig.2(a) of the paper was fabricated as follows. A 1145 nm thick layer of SiO\textsubscript{2} was grown on a single-side polished, cleaned Si wafer via wet-chlorinated thermal oxide growth in a furnace at 1000 $^{\circ}$C. A 125 nm thick stochiometric Si\textsubscript{3}N\textsubscript{4} layer was then deposited in a low-pressure chemical vapor deposition furnace. Layer thicknesses and indices were measured using an ellipsometer. Extracted refractive indices of the SiO\textsubscript{2} and Si\textsubscript{3}N\textsubscript{4} layers at 632 nm were 1.46 and 2.00, respectively. PMMA photoresist was used with electron-beam lithography to define and etch the second-order grating structure in the nitride layer. The grating period $\Lambda$ was chosen to equal $\lambda/n_{eff}$, where $\lambda$ is the wavelength corresponding to the exciton-trion resonances and $n_{eff}$ is the effective index of the guided mode in the nitride layer. The grating trench had a width of 80 nm and a depth of 90 nm. In order to improve the contrast in reflection-based polariton dispersion measurements discussed below, care was taken to ensure that the incident light did not directly couple to the MoSe\textsubscript{2} ML. To do so, thickness of the SiO\textsubscript{2} cladding layer was chosen such that interference between the incident and reflected light (from the Si-SiO\textsubscript{2} interface) results in a field intensity minimum near the location of the MoSe\textsubscript{2} ML, while maintaining the coupling efficiency of incident light into the waveguide mode high enough to allow measurements of polariton dispersion. The low contrast in the reflection spectrum shown in Fig.2(d) is due to the minimum in the light intensity at the location of the ML, as discussed above. Bulk MoSe\textsubscript{2} crystals, obtained from {\em 2D Semiconductors Inc.}, were exfoliated and transferred onto a PDMS film bonded to a glass slide. Monolayers on the PDMS film were identified by optical microscopy, and a custom-built transfer setup was used to place the monolayers onto the grating devices.

\subsection*{\label{sec:level11} Optical Measurements Setup}
Energy and momentum resolved reflection and PL spectroscopies were performed using a Fourier microscopy setup. The microscopy setup included a 0.6 numerical aperture (NA) objective, followed by a spatial filter that collected light from a sample area of 12 $\mu$m diameter, and a spectrometer. The input slit of the spectrometer served as a k-space aperture filter that selected light with nearly zero momentum in the x-direction (i.e. light with $k_{x} \approx 0$). The resulting energy and momentum ($k_{z}$) dispersed spectrum inside the spectrometer was focused onto the image plane of a cooled $1024 \times 256$ pixel charge-coupled-device (CCD) detector. In reflection spectroscopy, broadband light from a quartz-tungsten lamp was used to illuminate the sample. In PL spectroscopy, a 532 nm pump laser was used to excite the sample. The maximum pump intensity used on the sample was $\sim$100 $\mu$W/$\mu$m$^2$. The sample was kept at cryogenic temperatures using a liquid helium exchange cryostat.

\subsection*{\label{sec:level12} Computation of the Polariton Dispersion and Reflection Spectra}
Electromagnetic computations in this work were performed using the finite-difference frequency-domain (FDFD) technique~\cite{FDFD}. Refractive index values of 1.46, 2.00, and 3.7 were used for SiO\textsubscript{2}, Si\textsubscript{3}N\textsubscript{4}, and Si, respectively. The energy band dispersion of the bare PC waveguide (without the MoSe\textsubscript{2} ML on top) was calculated using the FDFD technique in COMSOL. The results are shown in Fig.2(b).

The contribution to the photon self-energy $\Sigma^{ph}(\vec{k},\omega)$ from excitons/trions can be expressed in terms of the 2D optical conductivity $\sigma(\vec{k},\omega)$ of the MoSe\textsubscript{2} ML,
\begin{equation}
\Sigma^{ph}(\vec{k},\omega) = -i\hbar \frac{|\chi(z^*)|^{2}}{2 \langle  \epsilon \rangle} \sigma(\vec{k},\omega)
\end{equation}
Here, $\chi(z^*)$ is the waveguide mode amplitude at the location $z^*$ of the ML and $\langle  \epsilon \rangle$ is the average dielectric constant experienced by the waveguide mode. Therefore, the MoSe\textsubscript{2} ML in FDFD simulations was modeled as a layer of thickness $d = 0.7$ nm~\cite{mose2thickness} with a 3D optical conductivity $\sigma_{\text{total}}(\vec{k},\omega)$ given by, $\sigma_{\text{total}}(\vec{k},\omega) = \sigma(\vec{k},\omega)/d - i \epsilon \omega$. Here, $\epsilon = 20 \epsilon_{o}$ is the high-frequency dielectric constant of MoSe\textsubscript{2} resulting from optical transitions at energies higher than the exciton-trion energies~\cite{mose2permittivity}. The expression for the 2D optical conductivity $\sigma(\vec{k},\omega)$ of the MoSe\textsubscript{2} ML is given by Rana et al.~\cite{ranamanybodytrion}. $\sigma(\vec{k},\omega)$ was computed using the experimentally determined values of the sample electron density and the exciton-trion linewidths, as only these quantities are needed to compute $\sigma(\vec{k},\omega)$~\cite{ranamanybodytrion}. The real part of the computed optical conductivity is shown in Fig.2(d)(top right plot). The optical conductivity thus obtained was used to compute the polariton energy band dispersion using FDFD and the results are shown as dashed lines in Fig.4. The bare PC waveguide reflection spectrum (shown in Fig.3) and the polariton reflection spectrum (shown in Fig.4) were calculated as follows. For each pair $(\vec{k},\omega)$, an incident plane wave was excited at a planar port located above the device and the reflection coefficient was extracted as the scattering matrix element evaluated at the same port. The reflection spectra thus obtained was normalized to the reflection spectrum obtained from a similar device but which had no MoSe\textsubscript{2} ML and also no grating etched into the nitride layer.

\subsection*{\label{sec:sub1} The Nature of Trions and Exciton-Trion-Polaritons: A Comparison of Different Models}
In this Section, we consider various descriptions of trion states, and related polariton states, that have been presented in the literature and describe their shortcomings. We then explain how the model used in this paper overcomes these shortcomings and provides a consistent picture of excitons, trions, and exciton-trion-polaritons in doped 2D semiconductors. The theoretical details regarding the model used in this paper can be found in several recent publications~\cite{ranamanybodytrion, ranaradlft,ranaetp}.

The models used in this work will be discussed in the light of the following Hamiltonian that describes electrons and holes in an electron-doped monolayer of a TMD, such as MoSe\textsubscript{2}, near the $K$ and $K'$ points in the Brillouin zone, interacting with each other and with an in-plane polarized optical mode of in the rotating wave approximation,
\begin{widetext}
	\begin{eqnarray}
	& H = & H_{e} + H_{ph} + H_{e-ph} \nonumber \\
	& H_{e}  = & \sum_{\vec{k},s} E_{c,s}(\vec{k}) c_{s}^{\dagger}(\vec{k})c_{s}(\vec{k}) + \sum_{\vec{k},s} E_{v,s}(\vec{k}) b_{s}^{\dagger}(\vec{k})b_{s}(\vec{k}) + \frac{1}{A}\sum_{\vec{q},\vec{k},\vec{k}',s,s'} U(q)  c_{s}^{\dagger}(\vec{k}+\vec{q})b_{s'}^{\dagger}(\vec{k}'-\vec{q})b_{s'}(\vec{k}')c_{s}(\vec{k}) \nonumber \\
	& &  +  \frac{1}{2A}\sum_{\vec{q},\vec{k},\vec{k}',s,s'} V(q)  c_{s}^{\dagger}(\vec{k}+\vec{q})c_{s'}^{\dagger}(\vec{k}'-\vec{q})c_{s'}(\vec{k}')c_{s}(\vec{k}) \nonumber \\
	& H_{ph}  = & \sum_{\vec{Q}} \hbar \omega(\vec{Q}) a^{\dagger}(\vec{Q})a(\vec{Q}) \nonumber \\
	& H_{e-ph}  = & \frac{1}{\sqrt{A}}\sum_{\vec{Q},\vec{k},s} \left( g_{s}c_{s}^{\dagger}(\vec{k}+\vec{Q})b_{s}(\vec{k})a(\vec{Q})  + h.c \right)
	\label{eq:SH}
	\end{eqnarray}
\end{widetext}
Here, $E_{c,s}(\vec{k})$ and $E_{v,s}(\vec{k})$ are the conduction band (CB) and valence band (VB) energies. $c_{s}(\vec{k})$, $b_{s}(\vec{k})$, and $a(\vec{Q})$ are the destruction operators for electron in the CB, electrons in the VB, and photons in the waveguide mode. $s,s'$ represent the spin/valley  degrees of freedom in the 2D material. $s=\{\sigma,\tau\}$, where $\sigma = \pm 1$ and $\tau = \pm 1$ represent spin and valley degrees of freedom, respectively. We assume for simplicity that the electron and hole effective masses, $m_{e}$ and $m_{h}$, respectively, are independent of $s$. $U(\vec{q})$  represents Coulomb interaction between electrons in the CB and VB and $V(\vec{q})$ represents Coulomb interaction among the electrons in the CB. $\hbar \omega(\vec{Q})$ is the energy of a photon with in-plane momentum $\vec{Q}$, and $g_{s}$ is the electron-photon coupling constant. $g_{s}$ is assumed to be non-zero only for the case of the optical coupling between the top most valence band and the conduction band of the same spin (for $s=\{+1,+1\}$ or $s=\{-1,-1\}$).

In what follows, we will need to refer to exciton states. A standard exciton state with momentum $\vec{Q}$ can be written as,
\begin{equation}
| \psi^{ex}(\vec{Q}) \rangle = \frac{1}{\sqrt{A}} \sum_{k} \frac{\phi^{ex*}_{\vec{Q}}(\vec{k})}{N_{ex}}  c^{\dagger}_{s}(\vec{k}+\lambda_{e}\vec{Q}) b_{s}(\vec{k}-\lambda_{h}\vec{Q}) |GS \rangle 
\end{equation}
$A$ is the sample area. The normalization factor $N_{ex}$ equals $\sqrt{1 - f_{c,s}(\vec{k}+\lambda_{e}\vec{Q})}$ and $\lambda_{e} = 1 - \lambda_{h} = m_{e}/(m_{e} + m_{h})$. $|GS\rangle$ is the ground state of the electron-doped TMD monolayer consisting of a full VB and a partially filled CB. Electron occupation probability of states in the CB is given by the function $f_{c,s}(\vec{k})$.

Various trion states have been considered in the literature. We discuss them below and point out why they fail to describe coherent polaritons involving trions. \\
	
	1. Suppose one considers the following fermionic 3-body trion state,
	\begin{eqnarray}
	| \chi^{tr}(\vec{Q}) \rangle  &=& \frac{1}{A} \sum_{\vec{k}_{1},\vec{k}_{2}} \frac{\phi^{tr*}_{\vec{Q}}(\vec{k}_{1},s;\vec{k}_{2},s')}{N_{tr}}\times \nonumber \\ &&c^{\dagger}_{s}(\vec{\underline{k}}_{1}) c^{\dagger}_{s'}(\vec{\underline{k}}_{2}) b_{s}(\vec{\underline{k}}_{1} + \vec{\underline{k}}_{2} - \vec{Q}) |GS \rangle 
	\end{eqnarray}
	The normalization factor $N_{tr}$ equals $\sqrt{[ 1 - f_{c,s}(\vec{\underline{k}}_{1})][1 - f_{c,s'}(\vec{\underline{k}}_{2})]}$ and $\vec{\underline{k}}$ (underlined) stands for $\vec{k} + \xi \vec{Q}$, where $\xi = m_{e}/(2m_{e} + m_{h})$. The above 3-body trion state has no optical matrix element with the state $| GS \rangle \otimes | n_{\vec{Q}}=1 \rangle$ (material in the ground state and one photon with momentum $\vec{Q}$ in the waveguide mode), i.e.,
	\begin{equation}
	\langle GS| \otimes \langle n_{\vec{Q}}=1 | H_{e-ph} |\chi^{tr}(\vec{Q}) \rangle = 0
	\end{equation}
	Therefore, this state cannot describe trions, nor can it be a part of coherent polaritons. \\

	2. Now suppose one considers the following 4-body trion state~\cite{EsserTrion,Ctrion,Crecent,Ctrion2, Ctrion3},
	\begin{eqnarray}
	| \kappa^{tr}(\vec{Q},\vec{p}) \rangle &=&    \frac{1}{A} \sum_{\vec{k}_{1},\vec{k}_{2}} \frac{\phi^{tr*}_{\vec{Q}}(\vec{k}_{1},s;\vec{k}_{2},s')}{N_{tr}}\times \nonumber \\ &&c^{\dagger}_{s}(\vec{\underline{k}}_{1}) c^{\dagger}_{s'}(\vec{\underline{k}}_{2}) b_{s}(\vec{\underline{k}}_{1} + \vec{\underline{k}}_{2}-(\vec{Q}+\vec{p})) c_{s'}(\vec{p}) |GS \rangle \nonumber \\
	\end{eqnarray}
	The above state includes a hole left behind in the Fermi sea when an electron with momentum $\vec{p}$ was scattered out of the Fermi sea by the photogenerated exciton to form a trion~\cite{EsserTrion,Ctrion,Crecent,Ctrion2,Ctrion3}. Here, $N_{tr}$ equals $\sqrt{[1+\delta_{s,s'}][1 - f_{c,s}(\vec{\underline{k}}_{1})][1 - f_{c,s'}(\vec{\underline{k}}_{2})]f_{c,s'}(\vec{p})}$ and $\vec{\underline{k}}$ (underlined) stands for $\vec{k} + \xi (\vec{Q} + \vec{p})$. The above state has a non-zero optical matrix element with the state $| GS \rangle \otimes | n_{\vec{Q}}=1 \rangle$, but this matrix element scales as $\sim 1/\sqrt{A}$ and is rather small ($A$ is the  sample area). In addition, the above trion state also has equally small optical matrix elements with all other states of the form $c_{s'}^{\dagger}(\vec{Q}-\vec{Q'} + \vec{p})c_{s'}(\vec{p})| GS \rangle \otimes |n_{\vec{Q'}}=1 \rangle$ consisting of a photon with momentum $\vec{Q'} \ne \vec{Q}$ in the waveguide mode and the material in the excited state with one electron outside the Fermi sea with momentum $\vec{Q}-\vec{Q'} + \vec{p}$ and one hole inside the Fermi sea with momentum $-\vec{p}$. What this means is that the state created by abosrbing a photon of momentum $\vec{Q}$, can decay by emitting a photon of a different momentum $\vec{Q'}$. Consequently, the above state is also not a suitable state for coherent polaritons involving trions.\\
	
	3. The smallness of the optical matrix elements found in the case considered above can be fixed by making the Fermi hole surround the trion. This leads to the following bosonic 4-body trion state~\cite{Emmanuele2020,Nonlineartr20,Glazov},
	\begin{eqnarray}
	| \beta^{tr}(\vec{Q}) \rangle &=&    \frac{1}{\sqrt{A^{3}}} \sum_{\vec{k}_{1},\vec{k}_{2},\vec{p}} \frac{\phi^{tr*}_{\vec{Q}}(\vec{k}_{1},s;\vec{k}_{2},s';\vec{p},s')}{N_{tr}}\times \nonumber \, \\ &&c^{\dagger}_{s}(\vec{\underline{k}}_{1}) c^{\dagger}_{s'}(\vec{\underline{k}}_{2}) b_{s}(\vec{\underline{k}}_{1} + \vec{\underline{k}}_{2}-(\vec{Q}+\vec{p})) c_{s'}(\vec{p}) |GS \rangle \nonumber\\
	\end{eqnarray}
	$N_{tr}$ here the same value as in case (2) considered above and $\vec{\underline{k}}$ (underlined) stands for $\vec{k} + \xi (\vec{Q} + \vec{p})$. The above 4-body trion state has a large non-zero optical matrix element with the state $| GS \rangle \otimes | n_{\vec{Q}}=1 \rangle$, and this matrix element can be shown to be proportional to $\sqrt{n_{s'}}$ where the electron density $n_{s'} = \sum_{\vec{p}} f_{c,s'}(\vec{p})/A$. Similar 4-body states have been proposed by Emmanuele {\em et al.}~\cite{Emmanuele2020}, Kyriienko {\em et al.}~\cite{Nonlineartr20}, and Glazov~\cite{Glazov}. It is then tempting to set up the following $3\times 3$ matrix for exciton-trion-polaritons using as basis the photon state $| GS \rangle \otimes | n_{\vec{Q}}=1 \rangle$, the exciton state $| \psi^{ex}(\vec{Q}) \rangle$, and the 4-body trion state $| \beta^{tr}(\vec{Q}) \rangle$~\cite{Emmanuele2020,Cuadra2018,Nonlineartr20,Dufferwiel2017},   
	\begin{eqnarray}
	&\begin{bmatrix}
	\hbar \omega (\vec{Q}) - i\gamma_{ph} &  \frac{1}{2}\hbar \Omega^{ex}  &  \frac{1}{2}\hbar \Omega^{tr} \\
	\frac{1}{2}\hbar \Omega^{ex} & E^{ex}(\vec{Q}) -i\gamma_{ex} & 0 \\
	\frac{1}{2}\hbar \Omega^{tr} & 0 & E^{tr}(\vec{Q}) - i\gamma_{tr} 
	\end{bmatrix} \nonumber \\
	&\label{eq:Seq2}
	\end{eqnarray}
	Here, $\hbar \Omega^{ex}/2$ and $\hbar \Omega^{tr}/2$ are the optical matrix elements of the exciton and trion states with the state $| GS \rangle \otimes | n_{\vec{Q}}=1 \rangle$, respectively. Similar $3\times 3$ matrices have been used by several authors, including Emmanuele {\em et al.}~\cite{Emmanuele2020}, Kyriienko {\em et al.}~\cite{Nonlineartr20}, Dufferwiel {\em et al.}~\cite{Dufferwiel2017}, Cuadra {\em et al.}~\cite{Cuadra2018}, and Rapaport {\em et al.}~\cite{SPfeiffer01}, to describe polaritons. The problem with the above matrix is that the 4-body trion state used in this model is not orthogonal to the exciton state used in the model and their inner product is proportional to $\sqrt{n_{s'}}$. This may not seem like a big problem and one may include this non-zero overlap in the variational scheme or, equivalently, use the standard linear algebra technique of Gram-Schmidt and orthogonalize the 4-body trion state $|\beta^{tr}(\vec{Q}) \rangle$ with respect to the exciton state (In carrying out this orthogonalization procedure one has to orthogonalize the trion state $|\beta^{tr}(\vec{Q}) \rangle$ with respect to all the exciton states.) However, the largest overlap of the lowest energy bound trion state $|\beta^{tr}(\vec{Q}) \rangle$ is with the lowest energy exciton state. But when this process is carried out, the resulting 4-body trion state turns out to have zero optical matrix element with the state $| GS \rangle \otimes | n_{\vec{Q}}=1 \rangle$. In other words, the new $3\times 3$ matrix after this orthogonalization procedure becomes,
	\begin{eqnarray}
	&\begin{bmatrix}
	\hbar \omega(\vec{Q})-i\gamma_{ph} &  \frac{1}{2}\hbar \Omega^{ex}  & 0 \\
	\frac{1}{2}\hbar \Omega^{ex} & E^{ex}(\vec{Q}-i\gamma_{ex} & 0 \\
	0 & 0 & E^{tr}_{new}(\vec{Q}) - i\gamma_{tr} 
	\end{bmatrix} \nonumber \\
	&\label{eq:Seq3}
	\end{eqnarray}
	Consequently, coherent exciton-trion-polaritons are not possible in this model, only exciton-poalritons are possible. This fact has been overlooked in many published works~\cite{Emmanuele2020,Cuadra2018,Nonlineartr20,Dufferwiel2017}. The failure of the state $| \beta^{tr}(\vec{Q}) \rangle$ in capturing the trion and polariton physics can also be demonstrated in several other ways. For example, the optical conductivity of the material is supposed to satisfy the Thomas-Reiche-Kuhn sum rule~\cite{ranamanybodytrion}. The optical conductivity that comes out from the matrix model in Eq.~\ref{eq:Seq2} violates this sum rule. This violation happens in the following way. As the electron density increases, and the trion state optical matrix element $\hbar \Omega^{tr}/2$ increases as $\sqrt{n_{s'}}$, the exciton state optical matrix element $\hbar \Omega^{ex}/2$ does not decrease but remains constant and the sum rule gets violated. The violation of the sum rule points to the fact that the exciton and trion states are not good approximate eigenstates of the Hamiltonian $H_{e}$ in Eq.~\ref{eq:SH}.


We now present our model for exciton-trion-polaritons in its simplest form. Theoretical details can be found in recent publications~\cite{ranamanybodytrion, ranaradlft,ranaetp}. An important piece of physics missing from the model considered above in case (3) is that in doped materials, exciton and trion states are not eigenstates of the Hamiltonian $H_{e}$ in Eq.~\ref{eq:SH}. Exciton and trion states are strongly coupled as a result of electron-electron and electron-hole Coulomb interactions. The many-body density matrix technique used by us resulted in a description of exciton-trion-polaritons in the form of a coupled system of a 2-body exciton state, 4-body bound and unbound trion states, and a photon state~\cite{ranamanybodytrion,ranaetp}. The 4-body bound trion state that results from the many body density matrix technique is~\cite{ranamanybodytrion},
\begin{eqnarray}
| \psi^{tr}(\vec{Q}) \rangle &=& \frac{1}{\sqrt{A^{3}}} \sum^{\vec{\underline{k}}_{1},\vec{\underline{k}}_{2} \ne \vec{p}}_{\vec{k}_{1},\vec{k}_{2},\vec{p}} \frac{\phi^{tr*}_{\vec{Q}}(\vec{k}_{1},s;\vec{k}_{2},s';\vec{p},s')}{N_{tr}}\times \nonumber \\
 &&c^{\dagger}_{s}(\vec{\underline{k}}_{1}) c^{\dagger}_{s'}(\vec{\underline{k}}_{2}) b_{s}(\vec{\underline{k}}_{1} + \vec{\underline{k}}_{2}-(\vec{Q}+\vec{p})) c_{s'}(\vec{p}) |GS \rangle \nonumber \\
\end{eqnarray}
$N_{tr}$ again has the same value as in case (2) considered above and $\vec{\underline{k}}$ (underlined) stands for $\vec{k} + \xi (\vec{Q} + \vec{p})$. The above trion state is orthogonal to all the exciton states, and also has a zero optical matrix element with the state $| GS \rangle \otimes | n_{\vec{Q}}=1 \rangle$. It is an eigenstate of a 4-body Schrodinger equation~\cite{ranamanybodytrion} but it is not an eigenstate of the Hamiltonian $H_{e}$. One needs to include Coulomb interactions between this trion state and the exciton state. When these Coulomb interactions, as well as interaction with light, are included, the resulting $3\times 3$ matrix for exciton-trion-polaritons is found to be,
\begin{eqnarray}
&\begin{bmatrix}
\hbar \omega(\vec{Q})- i\gamma_{ph} &  \frac{1}{2}\hbar \Omega^{ex}  & 0 \\
\frac{1}{2}\hbar \Omega^{ex} & E^{ex}(\vec{Q})-i\gamma_{ex} &  M_{b} \\
0 &  M_{b} & E^{tr}_{b}(\vec{Q}) - i\gamma_{tr} 
\end{bmatrix} \nonumber \\
&\label{eq:Seq4}
\end{eqnarray}  
$M_{b}$ is a Coulomb matrix element between the exciton and the 4-body bound trion states and its expression is given by Rana {\em et al.}~\cite{ranamanybodytrion}. $M_{b}$ is an increasing function of the electron density. There is still an important ingredient missing from the above matrix model and we discuss this missing piece next. In the absence of light, exciton and bound trion superposition states can be constructed and these states can diagonalize the lower $2 \times 2$ matrix of the above $3\times 3$ matrix. However, these exciton and bound trion superposition states are still not good eigenstates of the Hamiltonian $H_{e}$. So, for example, the energy splitting obtained between the two energy eigenvalues of the lower $2 \times 2$ matrix turns out to be much larger than the experimentally observed splitting between the two peaks in the optical absorption spectra of 2D materials~\cite{ranamanybodytrion}. What has been ignored in the above $3\times 3$ matrix is the continuum of unbound trion states (or exciton-electron scattering states) whose energies lie just above the exciton energy. These unbound trion states are also eigenstates of the 4-body Schrodinger equation~\cite{ranamanybodytrion} and are orthogonal to the exciton states. This continuum, when included in the analysis, prevents the upper energy eigenvalue of the lower $2 \times 2$ matrix from varying too much as the electron density increases and the resulting energy splitting then goes almost linearly with the Fermi energy, in agreement with the experiments~\cite{ranamanybodytrion}. This continuum of unbound trion states represents the screening of the exciton by the electrons in the electron-doped material. Once the continuum of unbound trion states is included in the analysis, the matrix for exciton-trion-polaritons becomes,
\begin{widetext}
	\begin{equation}
	\begin{bmatrix}
	\hbar \omega(\vec{Q})-i\gamma_{ph} &  \frac{1}{2}\hbar \Omega^{ex}  &  0 & 0 & 0 & 0 & \cdots \\
	\frac{1}{2}\hbar \Omega^{ex} & E^{ex}(\vec{Q})-i\gamma_{ex} & M_{b} & M_{1,ub} & M_{2,ub} & M_{3,ub} & \cdots \\
	0 & M_{b} & E^{tr}_{b}(\vec{Q}) - i\gamma_{tr} & 0 & 0 & 0 & \cdots \\
	0 & M_{1,ub} & 0 & E^{tr}_{1,ub}(\vec{Q}) - i\gamma_{tr} & 0 & 0 & \cdots \\
	0 & M_{2,ub} & 0 & 0 & E^{tr}_{2,ub}(\vec{Q}) - i\gamma_{tr} & 0 & \cdots \\
	0 & M_{3,ub} & 0 & 0 & 0 & E^{tr}_{3,ub}(\vec{Q}) - i\gamma_{tr} & \cdots \\
	\vdots & \vdots & \vdots & \vdots & \vdots & \vdots & \ddots
	\end{bmatrix}
	\label{eq:eq5}
	\end{equation}
\end{widetext}
The inclusion of unbound trion states not only modifies the energies of the two peaks observed in the optical spectra of 2D materials but also modifies their respective peak optical oscillator strengths~\cite{ranamanybodytrion}. The (infinite) matrix shown above cannot be reduced to a $3\times 3$ matrix in any meaningful way. The structure of exciton-trion-polaritons described by the matrix above is essentially the same as that expressed in the language of many body Green's functions (and optical conductivity) in the manuscript. It should also be noted here that the coupled Schr{\"o}dinger equation model of Rana {\em et al.}~\cite{ranamanybodytrion}, which is represented by the inifinite matrix model above, is not expected to be accurate at very large doping densities ($> 2\times10^{13}$ cm$^{-2}$ for most 2D TMDs) at which trion-electron interactions cannot be ignored~\cite{ranamanybodytrion}. At very large electron densities, multiple (and not just single) electron-hole excitations across the Fermi sea by the excitons need to be included in the analysis~\cite{ranamanybodytrion}.

For the sake of clarity, spin and valley indices, relevant to a MoSe$_{2}$ monolayer, have been suppressed in writing the above matrix. It should be noted that the exciton and trion states represented in the matrix are the transverse exciton and trion states. Transverse exciton states in TMD monolayers are superposition of exciton states from $K$ and $K'$ valleys that couple only to TE-polarized (in-plane polarized) light and do not couple to TM-polarized light. We have also assumed only one transverse exciton state, the one with the lowest energy, and this is a good approximation given the large energy separation between the lowest and the higher energy energy exciton states in 2D TMD monolayers. Finally, the above analysis can be easily extended to 2D TMDs such as WS$_{2}$ and WSe$_{2}$, in which the lowest conduction band has the opposite spin as that of the highest valence band in the same valley, provided both intravalley (spin singlet) and intervalley (spin triplet) trions, and their exchange splittings, are included in the analysis~\cite{Korn16}.

\subsection*{A Word on Exciton-Polarons}
The behavior of excitons in doped semiconductors was recently described in terms of polaron physics by Sidler {\em et al.}~\cite{atactrion}. Polaron physics associated with impurity atoms in cold Fermi gases has been extensively explored in the last two decades~\cite{demreview}. The screening of an impurity atom by the host Fermi atoms results in two eigenstates; the attractive and the repulsive polarons. A third solution, consisting of bound molecular state of the impurity atom and a host atom, can also exist~\cite{demreview}. The attractive and the repulsive polaron solutions are similar to the eigenstates expressed in terms of exciton-trion superposition states described by Rana {\em et al.}~\cite{ranamanybodytrion}. The exciton-trion superposition states are good approximate eigenstates of the Hamiltonian $H_{e}$ and can be written as~\cite{ranamanybodytrion},
\begin{widetext} 
	\begin{eqnarray}
	|\psi_{n,s}(\vec{Q})\rangle & = & \frac{\alpha}{\sqrt{A}} \sum_{k} \frac{\phi^{ex*}_{n,\vec{Q}}(\vec{k})}{N_{ex}} c^{\dagger}_{s}(\vec{k}+\lambda_{e}\vec{Q}) b_{s}(\vec{k}-\lambda_{h}\vec{Q}) |GS \rangle \nonumber \\
	&& + \sum_{m,s'} \frac{\beta_{m,s'}}{\sqrt{A^{3}}} \sum^{\vec{\underline{k}}_{1},\vec{\underline{k}}_{2} \ne \vec{p}}_{\vec{k}_{1},\vec{k}_{2},\vec{p}} \frac{\phi^{tr*}_{m,\vec{Q}}(\vec{k}_{1},s;\vec{k}_{2},s';\vec{p},s')}{N_{tr}} c^{\dagger}_{s}(\vec{\underline{k}}_{1}) c^{\dagger}_{s'}(\vec{\underline{k}}_{2}) b_{s}(\vec{\underline{k}}_{1} + \vec{\underline{k}}_{2}-(\vec{Q}+\vec{p})) c_{s'}(\vec{p}) |GS \rangle \\
	\label{eq:var} \nonumber
	\end{eqnarray}
\end{widetext}
The exciton state is labeled by the subscript $n$. The solution involves  a summation over all bound and unbound trion states labeled by the subscript $m$, consistent with the values of $s$ and $s'$. The above state is very similar to the Chevy's ansatz for attractive and repulsive Fermi polaron states~\cite{Chevy06,atactrion}. A difference is that in the Chevy's ansatz the trion state is not taken to be orthogonal to the exciton state leading to incorrect optical matrix elements. The bound molecular state, on the other hand, is essentially the same as the 3-body fermionic trion state considered above (in case 1),
\begin{eqnarray}
| \chi^{tr}(\vec{Q}) \rangle & = &\frac{1}{A} \sum_{\vec{k}_{1},\vec{k}_{2}} \frac{\phi^{tr*}_{\vec{Q}}(\vec{k}_{1},s;\vec{k}_{2},s')}{N_{tr}}\times \nonumber \\ &&c^{\dagger}_{s}(\vec{\underline{k}}_{1}) c^{\dagger}_{s'}(\vec{\underline{k}}_{2}) b_{s}(\vec{\underline{k}}_{1} + \vec{\underline{k}}_{2} - \vec{Q}) |GS \rangle
\end{eqnarray}
Sidler {\em et al.} had considered the trion state to be the same as the bound molecular state given above~\cite{atactrion}. In addition, Sidler {\em et al.} had claimed to see separate experimental signatures of both the bound molecular state and the attractive and repulsive polaron states in the measured optical spectra. While a 3-body fermionic state can in principle exist in the limit of very small electron densities, and can exhibit signatures in photoluminescence spectrum (but not in  the optical absorption spectrum), it is not a good eigenstate at the large ($>10^{12}$ cm$^{-2}$) electron densities considered in the work of Sidler {\em et al.}. It is likely that the photoluminescence attributed to molecular trion states in the work of Sidler {\em et al.} was due to localized states. Some care is needed when transporting results from the Fermi polaron literature involving atomic gases to excitons in doped semiconductors. In contrast to the short-range interactions in neutral Fermi atomic gases, the Coulomb interactions in 2D materials are long-range. A charged 3-body bound fermionic state cannot exist as such in a doped 2D material with a large electron density. It will immediately get surrounded by a screening hole and form an attractive polaron (or an exciton-trion superposition state). For the same reason, the trion plus hole continuum (states described in case 2 above), also found by Sidler {\em et al.} in their computations, are not possible in the presence of long-range interactions. Efimkin {\em et al.} identified the two peaks that are observed in the optical spectra of doped 2D materials as the attractive and repulsive polaron states~\cite{Efimkin17}. Their identification agrees with our identification of these peaks as the lower and higher energy exciton-trion superposition states. However, Efimkin {\em et al.} also used short-range potentials in their analysis and found the trion plus hole continuum (case 2 above), which are not physical states when long-range Coulomb potentials are used and when the interactions involving the Fermi hole are not ignored.

\end{document}